\documentclass[3p]{elsarticle}

\usepackage{graphicx}
\usepackage{amsfonts,amsmath}
\usepackage{multirow}

\usepackage[usernames,dvipsnames]{xcolor}

\begin{document}
	\journal{Submitted to R. Proc Soc A}

\begin{frontmatter}

    \title{A diffuse interface model 
	for the analysis of propagating bulges in cylindrical balloons}
	
    \author{C.\ Lestringant$^{1}$ and B.\ Audoly$^{2,3}$}
	
	\address{$^{1}$Department of Mechanical and Process Engineering, 
	ETH Z{\"u}rich, 8092 Z{\"u}rich, Switzerland\\
	$^{2}$Laboratoire de M\'ecanique des Solides, \'Ecole 
	Polytechnique and CNRS, F-91128 Palaiseau, France\\
	$^{3}$Division of Applied Science and Engineering, California
	Institute of Technology, Pasadena
	, USA}


    \begin{abstract}
		With the aim to characterize the formation and propagation of
		bulges in cylindrical rubber balloons, we carry out an
		expansion of the non-linear axisymmetric membrane model
		assuming slow axial variations.  We obtain a diffuse
		interface model similar to that introduced by van der Waals in
		the context of liquid-vapor phase transitions.  This provides
		a quantitative basis to the well-known analogy between
		propagating bulges and phase transitions.  The diffuse
		interface model is amenable to numerical as well as analytical
		solutions, including linear and non-linear bifurcation
		analyses.  Comparisons to the original membrane model reveal
		that the diffuse interface model captures the bulging
		phenomenon very accurately, even for well-localized phase
		boundaries.
    \end{abstract}

\end{frontmatter}

\section{Introduction}

\subsection{Problem and background}

Bulges in cylindrical rubber balloons are a classical example of
localization in solid mechanics.  When a balloon is inflated, an
initial regime of uniform inflation is followed by the formation of a
bulge: the bulge appears initially as a long-wavelength buckling mode
that localizes rapidly and then grows locally, until it propagates and
eventually invades the entire
balloon~\cite{Kyriakides-Chang-The-initiation-and-propagation-of-a-localized-1991}.
As in other localization phenomena, the formation of a bulge reflects
the non-convexity of the strain energy when restricted to homogeneous
deformations: the onset of bulging occurs quickly after the maximum in
the volume-pressure loading curve, and the propagation pressure can be
predicted by Maxwell's equal-area
rule~\cite{Chater-Hutchinson-On-the-propagation-of-bulges-and-buckles-1984}.
%
%



Several other localization phenomena have been studied in solid
mechanics, such as stress-induced phase transformations~\cite{%
Ericksen75,Bhattacharya-Oxford-Series-on-Materials-2004,%
FuFreidin04}, the necking of
bars~\cite{Bridgman-Studies-in-large-plastic-1952,%
Barenblatt-Neck-propagation-in-polymers-1974,%
Hutchinson-Miles-Bifurcation-analysis-of-the-onset-1974}, kink bands
in compressed fiber
composites~\cite{Wadee-Hunt-EtAl-Kink-band-instabilities-2004,FuZhang06},
as well as localized structures in thin elastic
shells~\cite{Power-Kyriakides-Localization-and-propagation-of-instabilities-1994}
and tape springs~\cite{%
Seffen-Pellegrino-Deployment-dynamics-of-tape-1999,%
Seffen-You-EtAl-Folding-and-deployment-of-curved-2000%
}.

These localization phenomena have been investigated based on two types
of models, as discussed
in~\cite{Triantafyllidis-Bardenhagen-The-influence-of-scale-size-1996}
for example.  On the one hand, \emph{non-regularized models}, also
known as sharp interface models, make use of a classical strain energy
functional depending solely on the strain: the onset of localization
is associated with the loss of ellipticity of the equations of
equilibrium at a critical value of the load~\cite{KnowlesSternberg}.
Such models can typically predict the critical load, the formation of
different phases and the orientation of the phase boundaries, but
cannot predict their subsequent evolution, nor their number or
distribution in space; they cannot resolve the displacement inside the
localized region either.  On the other hand, \emph{regularized
models}, also known as diffuse interface models, make use of a stored
elastic energy functional depending on both the strain and the
\emph{strain gradient}: such models remedy the limitations of the
non-regularized models, and in particular remain well posed beyond the
onset of localization~\cite{TriantaAifan}.

Regularized models are often introduced heuristically, but can in some
cases be justified mathematically.  Such a justification has been done
in the case of periodic elastic solids, such as elastic
crystals~\cite{Triantafyllidis-Bardenhagen-On-higher-order-gradient-1993,%
BardenTrianta94}, trusses made of elastic bars or
beams~\cite{Abdoul-Anziz-Seppecher-Strain-gradient-and-generalized-2018},
or elastic solids with a periodic
micro-structure~\cite{Triantafyllidis-Bardenhagen-The-influence-of-scale-size-1996,%
Bacigalupo-Paggi-EtAl-Identification-of-higher-order-continua-2017}.
In these works on periodic solids, the ratio $R/L\ll 1$ of microscopic
cell size $R$ to the macroscopic dimension $L$ of the structure is
used as an expansion parameter, and the homogenized properties of the
periodic medium are derived through a systematic expansion in terms of
the macroscopic strain and of its successive gradients.

The goal of this paper is to derive a one-dimensional, regularized
model applicable to the analysis of axisymmetric bulges in cylindrical
rubber balloons.  It is part of a general effort to characterize
localization phenomena occurring in slender structures, which have
been much less studied than in periodic solids.  In slender
structures, regularized models can be derived by an asymptotic
expansion as well, using now the aspect ratio $R/L \ll 1$ as the
expansion parameter where $R$ is the typical transverse dimension of
the structure and $L$ its length.  This approach has been carried out
for the analysis of necking, and diffuse interface models have been
derived asymptotically, first for a two-dimensional hyperelastic strip
by Mielke~\cite{Mielke-Hamiltonian-and-Lagrangian-flows-1991} and
later for a general prismatic solid in three dimensions by Audoly
\&~Hutchinson~\cite{BAJH}.  These authors proposed an expansion method
upon which we build ours (and which we improve).

Our asymptotic expansion starts from the axisymmetric membrane model,
which has been used extensively to analyze bulges in cylindrical
balloons~\cite{KyriaChang90,
Kyriakides-Chang-The-initiation-and-propagation-of-a-localized-1991}.
Its outcome is a one-dimensional diffuse interface model, exactly
similar to that introduced heuristically by van der
Waals~\cite{vdwaals94} to analyze the liquid-vapor phase transitions
at a mesoscopic level.  The analogy between bulges in balloons and
phase transitions has been known for a long time: Chater \&
Hutchinson~\cite{Chater-Hutchinson-On-the-propagation-of-bulges-and-buckles-1984}
have adapted Maxwell's rule for the the coexistence of two phases to
derive the pressure at which a bulge can propagate in a balloon, while
M\"uller \&
Strehlow~\cite{Muller-Strehlow-Rubber-and-Rubber-Balloons-2004} have
proposed a pedagogical introduction to the theory of phase transitions
based on the mechanics of rubber balloons.  Here, we push the analogy
further, and show that the diffuse interface model can provide a
quantitative description of bulges in balloons, not only accounting
for the propagation pressure, but also for the domain boundary between
the bulged and unbulged phases, as well as for its formation via a
bifurcation---borrowing from the theory of phase transitions, we will
refer to this boundary as a `diffuse interface'.  The diffuse
interface model is classical, tractable, and amenable to analytical
bifurcation and post-bifurcation analysis, as we demonstrate.  It is
also simpler than the axisymmetric membrane model on which it is
based.

There is a vast body of work on the bulging of cylindrical balloons,
all of which have used the theory of axisymmetric membranes as a
starting point.
The stability and bifurcations from homogeneous solutions have been
analyzed in~\cite{CorneliussenShield61,Shield72, HaughtonOgden79}.
Non-linear solutions comprising bulges have been derived
in~\cite{Yin-Non-uniform-inflation-of-a-cylindrical-1977}.  The
analysis of stability has been later extended to arbitrary
incompressible hyperelastic materials, to various closure conditions
at the ends of the tube, as well as to various type of loading
controls based on either the internal pressure, the mass or the volume
of the enclosed gas~\cite{chenbulges}.
In a recent series of four
papers, Fu \textit{et al.}~\cite{Fu-Pearce-EtAl-Post-bifurcation-analysis-of-a-thin-walled-2008,Fu-Xie-Stability-of-localized-bulging-2010,Pearce-Fu-Characterization-and-stability-of-localized-2010,Fu-Effects-of-imperfections-on-localized-2012}
revisit the bifurcation problem, complement it with the weakly non-linear post-bifurcation
analysis in the case of an infinite tube, and address
imperfection sensitivity.  
Besides these theoretical studies, there has been a number of
experimental and numerical papers on balloons.  A compelling agreement
between experiments and numerical simulations of the non-linear
membrane model has been obtained
by Kyriakides \& Chang~\cite{KyriaChang90,%
Kyriakides-Chang-The-initiation-and-propagation-of-a-localized-1991},
who provide detailed experimental and numerical results on the
initiation, growth and propagation of bulges, highlighting the analogy
with phase transitions.  Given that the agreement between experiments
and the non-linear membrane theory has already been covered thoroughly
in this work, our focus here will be on comparing the diffuse
interface model to the non-linear membrane model, using exactly the
same material model as in Kyriakides' simulations and experiments.

\subsection{Outline of the main results}

\begin{figure}
	\centering
	\includegraphics[scale=.90]{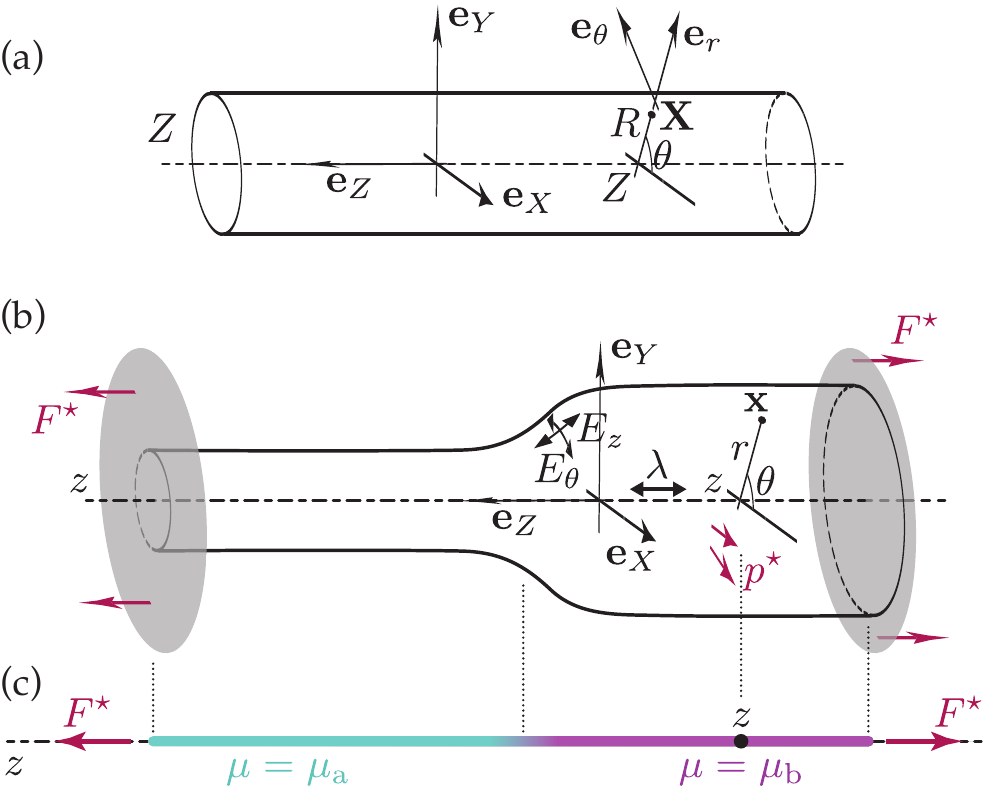}
	\caption{Inflation of a cylindrical membrane: (a)~reference
	configuration, (b)~sketch of an equilibrium configuration when the
	membrane is subject to an axial force $F$ and an internal pressure
	$p$.  (c) Equivalent diffuse interface model derived in this
	paper: a bar having an order parameter $\mu(Z)$ undergoes a phase
	transformation.}
	\label{fig:experiment}
\end{figure}

Our work focusses on solutions to the non-linear axisymmetric membrane
model that vary slowly in the axial direction, as happens typically at
the onset of localization.  A systematic expansion of the membrane
energy is obtained in terms of the aspect-ratio parameter
$\varepsilon=R/L\ll 1$, where $R$ is the initial radius of the balloon and
$L$ its initial length.  The result reads
\begin{equation}
	\mathcal{E} [p, \mu] = \int_{0}^{L} \left[ G_0 (p, \mu (Z)) + \frac{1}{2} B_0 (p,
	\mu (Z)) \mu^{\prime 2} (Z) \right] \mathrm{d} Z 
	\label{eq:diffuseInterfaceModel-Announce}
\end{equation}
where $p$ denotes the (scaled) internal pressure, a control parameter
in the experiments, $Z$ is the axial coordinate and $\mu (Z)$ is a
strain measure, see figure~\ref{fig:experiment}.  Specifically, $\mu$
is the orthoradial stretch, defined as the ratio $\mu(Z)= \frac{r
(Z)}{R}$ of the current radius $r$ to the initial radius.
The potential $G_{0}$ appearing in the first term is a non-convex
function of the stretch $\mu$, much like in Ericksen's
bar~\cite{Ericksen75}: the non-regularized model for the balloon would
correspond to the energy functional
$\int_{0}^{L}G_{0}(p,\mu(Z))\,\mathrm{d}Z$.  Values of $p$ such that
several minima of $G_0$ exist correspond to pressures for which
different phases (associated with different values of the stretch
$\mu$) are in competition.  The second term
$\frac{1}{2}\,B_{0}\,\mu'^{2}$ in the integrand is a correction of
order $\varepsilon^{2}$, that accounts for the energetic cost of
inhomogeneity; in the theory of phase transitions, this is the term
that would account for surface tension at an interface.

We provide simple and explicit formulas for both the potential $G_0$
characterizing homogeneous solutions, see \S\ref{s:homsol_barmod},
equations~(\ref{eq:strainEnergy}), (\ref{eq:H0Decomposition})
(\ref{eq:lambda0def}) and~(\ref{eq:Gdef}) in particular, and for the
modulus $B_0$ of the regularizing term, see \S\ref{s:reducedmodel} and
equation~(\ref{eq:Bexpression}).

The diffuse interface model is obtained by a systematic, formal
expansion.  It is asymptotically exact and does not rely on any
unjustified kinematic assumptions:
equation~(\ref{eq:diffuseInterfaceModel-Announce}) approximates the
energy of the original membrane model with an error of order
$\varepsilon^{4}$ that is negligible compared to the smallest term
retained, namely the gradient term of order $\varepsilon^{2}$.  By
contrast, regularized models for slender structures have been proposed
in earlier work starting from kinematic hypotheses, which appeared to
be incorrect: see the treatment of necking in an elastic cylinder
in~\cite{ColemanNewman} as well as the critical discussion
in~\cite{BAJH}.

Our derivation is based on a \emph{finite-strain} membrane model.  The
non-linear features of the elastic constitutive law at finite strain
are ultimately reflected in the diffuse interface model through the
non-linear potential $G_{0}(p,\mu)$ and through the dependence of the
second gradient coefficient $B_{0}(p,\mu)$ on the current strain
$\mu$.  By contrast, an assumption of small strain has been used in
previous work~\cite{CaiDai06P1,%
Dai-Wang-An-analytical-study-on-the-geometrical-2009} on the
justification of a diffuse interface model to analyze phase
transformations in an elastic cylinder: this assumption is
questionable since the presence of coexisting phases involves finite
variations of strain across the interface.


The outline of the paper is as follows.
In~\S\ref{s:nonlinearMembraneModel}, we introduce the non-linear
membrane model.  In~\S\ref{s:homsol_barmod} we analyze its homogeneous
solutions, and derive an expression for the potential $G_{0}$.
Section~\ref{s:reducedmodel} is the core of the paper, and establishes
the diffuse interface model~(\ref{eq:diffuseInterfaceModel-Announce})
by an asymptotic method.  Section~\ref{sec:application} derives
solutions to the diffuse interface model using various methods, and
compares them with the predictions of the original membrane model.

\section{Non-linear membrane model}
\label{s:nonlinearMembraneModel}

We consider a cylindrical membrane with uniform initial thickness $H$
and radius $R$.  We use the cylindrical coordinates $(Z,\theta)$ in
reference configuration as Lagrangian variables.  When subject to
external load, the cylinder deforms into an axisymmetric membrane, see
figure~\ref{fig:experiment}a.  The cylindrical coordinates of a
material point in actual configuration are written as $(r(Z),\theta)$,
corresponding to a position $\mathbf{x} (Z, \theta) = z (Z)
\,\mathbf{e}_z + r (Z)\,\mathbf{e}_r (\theta)$, where
$(\mathbf{e}_{r}(\theta), \mathbf{e}_{\theta}(\theta),
\mathbf{e}_{z})$ is the local cylindrical basis.

In the axisymmetric membrane theory, the deformation gradient is a
$3\times 2$ matrix which writes
\[ \mathbf{F}=
\mu\,\mathbf{e_{\theta}}\otimes \mathbf{e}_{\theta}
+
(R\,\mu'\,\mathbf{e}_{r} + \lambda\,\mathbf{e}_{z})\otimes 
\mathbf{e}_{z}
\]
where we have defined an apparent axial stretch $\lambda$ and the
circumferential stretch $\mu$ as
\begin{subequations}
\begin{align}
	\lambda(Z) & = \frac{\mathrm{d} z }{\mathrm{d} Z}(Z),\\
	\mu(Z) & = \frac{r (Z)}{R}.
\end{align}
\end{subequations}
The Green-Lagrange strain tensor $\mathbf{E}= \frac{1}{2}
(\mathbf{F}^T \cdot \mathbf{F}-\mathbf{1})$ is a $2\times 2$ diagonal
matrix in the basis $(\mathbf{e}_{\theta},\mathbf{e}_{z})$ tangent to
the undeformed mid-surface: it will be represented compactly as a
vector, whose entries are the diagonal components $E_{\theta}$ and
$E_{z}$ of the matrix,
\begin{equation}
\mathbf{E} (\lambda, \mu, \mu') = \left(\begin{array}{c}
E_{\theta}\\
E_z
\end{array}\right) 
= \frac{1}{2}  \begin{pmatrix}
\mu^2 - 1\\
\lambda^2 + (R \mu')^2 - 1
\end{pmatrix}, \label{eq:genericMembraneStrain}
\end{equation}
where $\mu' = \frac{\mathrm{d}\mu}{\mathrm{d}Z}$ is a stretch
gradient, namely the axial gradient of circumferential stretch.

A material model is now specified through a strain energy per unit
volume $w^{*}(E_{\theta},E_{z})$.  In previous work on axisymmetric membranes~\cite{KyriaChang90,%
Kyriakides-Chang-The-initiation-and-propagation-of-a-localized-1991,%
Fu-Pearce-EtAl-Post-bifurcation-analysis-of-a-thin-walled-2008}, the
2d material model proposed
by Ogden~\cite{Ogden-Large-deformation-isotropic-1972} for incompressible
rubber has been used:
\begin{equation}
w^{*}(E_{\theta}, E_z)=  \sum_{i = 1}^3
\frac{S_i}{\alpha_i} \left(\ell_{\theta}^{\alpha_i} +  \ell_z^{\alpha_i} +
\left( \frac{1}{\ell_{\theta} \ell_z} \right)^{\alpha_i} \right),
\quad
\textrm{with }
\begin{cases}
	\ell_{\theta} = \sqrt{2 E_{\theta} + 1}  & \textrm{(circumfer.\ 
	stretch)}\textrm{,} \\
	\ell_z = \sqrt{2 E_z + 1}& \textrm{(axial  stretch)}\textrm{.}
\end{cases}
\label{eq:ogdenlaw}
\end{equation} 
We use this model as well for our numerical examples, with the same
set of material parameters $\alpha_i$'s and $S_i$'s as used in
previous work, namely $S_1 = 617$ kPa, $S_2 = 1.86$kPa,
$S_3=-9.79$kPa, $\alpha_1 = 1.3$, $\alpha_2 = 5.08$ and $\alpha_3 =
-2$.  All our results can be easily adapted to a different
constitutive law.  For this constitutive law, the initial shear
modulus $S_{\mathrm{ini}}$ can be obtained as $S_{\mathrm{ini}} =
\sum_{i=1}^{3}\alpha_{i}\,S_{i}$. 

The domain $0\leq Z\leq L$ represents one half of a balloon comprising
a single bulge center at $Z=0$, with symmetry conditions
$\mu'=\lambda' = 0$ enforced at $Z=0$.  At the other endpoint $Z=L$,
we consider the ideal boundary conditions sketched in
figure~\ref{fig:experiment}b, whereby the terminal section of the
balloon is resting and freely sliding on a planar `plug'.  These
conditions would be difficult to achieve in experiments but they offer
the advantage of being compatible with a uniform expansion of the
membrane, which simplifies the analysis.  By contrast, actual
cylindrical balloons are typically closed up on their ends and cannot
be inflated in a homogeneous manner due to end effects; these end
effects could reproduced by employing different boundary conditions,
but we prefer to ignore them.  Note that Kyriakides \&
Chang\cite{KyriaChang90,
Kyriakides-Chang-The-initiation-and-propagation-of-a-localized-1991}
use a rigid plug condition on one end, $\mu(L) = 1$, which is not
realistic either.  Our boundary conditions, sketched in
figure~\ref{fig:experiment} and provided in explicit form
in~\S\ref{pre-ssec:fullSimulations}\ref{ssec:fullSimulations}, are
natural: the applicable equilibrium condition will emerge
automatically from the condition that the energy is stationary.

As in the experiments of Kyriakides \&
Chang~\cite{Kyriakides-Chang-The-initiation-and-propagation-of-a-localized-1991},
the membrane is subject to an interior pressure $p^{\ast}$ and to a
stretching force $F^{\ast}$ applied along the axis, see
figure~\ref{fig:experiment}.  The total potential energy reads
\begin{equation}
	\mathcal{E}_{\mathrm{memb}}^{\ast} =
	\int_{0}^{L}\Big(
	w^{*}(\mathbf{E})\,2\,\pi\,R\,H\,\mathrm{d}Z
	-
	\pi\,(R\,\mu)^{2}\,(\lambda\,\mathrm{d}Z)\,p^{\ast}
	-
	(\lambda\,\mathrm{d}Z)\,F^{*}
	\Big)
	\nonumber
\end{equation}
where $2\,\pi\,R\,H\,\mathrm{d}Z$ is the initial volume element,
$\pi\,(R\,\mu)^{2}\,(\lambda\,\mathrm{d}Z) = \pi\,r^{2}\,\mathrm{d}z$
is the current enclosed volume element, and $\lambda\,\mathrm{d}Z =
\mathrm{d}z$ is the current axial length element.

We introduce a rescaled energy, denoted without an asterisk as
$\mathcal{E}_{\mathrm{memb}} = \frac{\mathcal{E}_{\mathrm{memb}}^{\ast}}{
(2\pi\,R\,H)\,S_{\mathrm{ini}}}$:
\begin{equation}
	\mathcal{E}_{\mathrm{memb}}[p,\lambda,\mu]= \int_{0}^{L}\left( w
	(\mathbf{E}(\lambda,\mu,\mu')) - p \frac{e}{2}\, \lambda \, \mu^2 - F \lambda
	\right) \mathrm{d} Z
	\textrm{.}
	\label{eq:completeEnergy}
\end{equation}
The strain energy, the force and pressure have been rescaled as well,
as $w = \frac{w^{*}}{S_{\mathrm{ini}}}$, $F = \frac{F^{\ast}}{2 \pi
R H S_{\mathrm{ini}} }$ and $p=\frac{p^{\ast}}{S_{\mathrm{ini}}}$,
respectively, and $e = \frac{R}{H}$ is an initial aspect ratio.  In
our numerical examples, we use the same value $e = \frac{55}{16}$ as
in~\cite{Kyriakides-Chang-The-initiation-and-propagation-of-a-localized-1991}:
even though this balloon is relatively thick prior to deformation, the
non-linear membrane model has been checked to match the experimental
results accurately
in~\cite{Kyriakides-Chang-The-initiation-and-propagation-of-a-localized-1991}.
We also use the same value of the load
\begin{equation}
	F = 1.149
	\label{eq:forceF}
\end{equation}
as in these experiments.  The parameter $F$ will never be changed, and
we do not keep track of how the various quantities depend on $F$; the
argument $F$ will systematically be omitted in functions, as we did
already in the left hand side of~(\ref{eq:completeEnergy}).

The functions $\lambda(Z)$ and $\mu(Z)$ that make the
energy~(\ref{eq:completeEnergy}) stationary yield the axisymmetric
equilibra of the balloon.  These solutions are obtained by a numerical
method described in
section~\ref{pre-ssec:fullSimulations}\ref{ssec:fullSimulations},
and are plotted as the black curves in
figure~\ref{fig:fullnonlinear_solutions}, where they are used as a
reference.

\section{Analysis of homogeneous solutions}
\label{s:homsol_barmod}

Our general goal is to justify the diffuse interface model when
$\lambda(Z)$ and $\mu(Z)$ vary slowly as a function of $Z$.  In this
section, we start by considering the case where $\lambda$ and $\mu$ do
not depend on $Z$,
\begin{equation}
	\frac{\mathrm{d}\lambda}{\mathrm{d}Z} = 0,\qquad
	\frac{\mathrm{d}\mu}{\mathrm{d}Z} = 0
	\textrm{.}
\end{equation}
This corresponds to homogeneous solutions, \emph{i.e.}\ to solutions
with uniform inflation.  These homogeneous solutions are well known,
and are re-derived here for the sake of completeness.  A catalog of
such homogeneous solutions will be obtained, which plays a key role in
the subsequent derivation of the diffuse interface model.

\subsection{Kinematics of homogeneous solutions}

For homogeneous solutions, the gradient term $\mu'$
in~(\ref{eq:genericMembraneStrain}) vanishes and the membrane strain 
reads
\begin{equation}
	\mathbf{E}_0 (\lambda, \mu) 
	=
	\begin{pmatrix}
		E_{0}^{\theta} \\
		E_{0}^{z}
	\end{pmatrix}
	= 
	\frac{1}{2}
	\left(
	\begin{array}{c}
		\mu^2 - 1\\
		\lambda^2 - 1
	\end{array}
	\right).
	\label{eq:E0}
\end{equation}
All the quantities pertaining to homogeneous solutions are denoted
using a subscript `$0$'.  In the homogeneous case, 
the strain energy becomes
\begin{equation}
	w_0 (\lambda, \mu) = 
	\frac{1}{S_{\mathrm{ini}}} \sum_{i = 1}^3
	\frac{S_i}{\alpha_i} \left(\lambda^{\alpha_i} + \mu^{\alpha_i} +
	\left( \frac{1}{\lambda\, \mu} \right)^{\alpha_i} \right)
	\textrm{.}
	\label{eq:strainEnergy}
\end{equation}

Of particular importance will be the second Piola-Kirchhoff membrane
stress $\boldsymbol{\Sigma}_{0}$, defined as the gradient of the strain
energy with respect to the strain:
\begin{equation}
	\boldsymbol{\Sigma}_0 (\lambda, \mu) =
	\begin{pmatrix}
		\Sigma_{0}^{\theta} \\ \Sigma_{0}^{z}
	\end{pmatrix}
	=
	\begin{pmatrix}
		\frac{\partial w}{\partial E_{\theta}} \\
		\frac{\partial w}{\partial E_{z}}
	\end{pmatrix}_{\mathbf{E} = \mathbf{E}_{0}}
	=
	\begin{pmatrix}
		\frac{1}{\mu}  \frac{\partial w_{0}}{\partial \mu} (\lambda, 
		\mu)\\
		\frac{1}{\lambda}  \frac{\partial w_{0}}{\partial \lambda} (\lambda, 
		\mu)
	\end{pmatrix}\textrm{.}
	\label{eq:sigma0}
\end{equation}

\subsection{Equilibrium of homogeneous solutions}

In view of~(\ref{eq:completeEnergy}), the total potential energy of a
homogeneous solution per unit reference length is
\begin{equation}
g_0 (p, \lambda, \mu) = w_0 (\lambda, \mu) - p \frac{e}{2}  \,\lambda \, \mu^2 - F
\lambda. \label{eq:H0Decomposition}
\end{equation}
Given the load parameters $p$ (and $F$) the equilibrium values of
$\lambda$ and $\mu$ are found by the condition of equilibrium in
the axial and transverse directions,
\begin{subequations}
\begin{eqnarray}
	\frac{\partial g_0}{\partial \lambda} (p, \lambda, \mu) & = & 0, 
	\label{eq:g0EquiAxial}\\
	\frac{\partial g_0}{\partial \mu} (p, \lambda, \mu) & = & 0. 
	\label{eq:g0EquiTransv}
\end{eqnarray}
\end{subequations}
We leave the load $p$ left unspecified for the moment, and we view the
axial equilibrium~(\ref{eq:g0EquiAxial}) as an implicit equation for
$\lambda = \lambda_0 (p, \mu)$ in terms of $p$ and $\mu$: by
definition, $\lambda_0(p, \mu)$ is the solution to the implicit
equation
\begin{equation}
	\frac{\partial g_0}{\partial \lambda} (p, \lambda_0 (p, \mu), \mu) = 0.
	\label{eq:lambda0def}
\end{equation}
From now on, we will systematically eliminate $\lambda = \lambda_0(p,
\mu)$ in favor of the second unknown $\mu$.  Starting with the
potential $g_{0}$, we define a reduced potential $G_{0}$ as
\begin{equation}
G_0 (p, \mu) = g_0 (p, \lambda_0 (p, \mu), \mu)\textrm{,}
\label{eq:Gdef}
\end{equation}
as well as the stress $n_0$ dual to $\mu$,
\begin{subequations}
\begin{equation}
n_0 (p, \mu) = - \frac{\partial G_0}{\partial \mu} (p, \mu)
\textrm{.}
\label{eq:n0label}
\end{equation}
This $n_0 (p, \mu)$ can be interpreted as an imbalance of hoop stress;
it vanishes at equilibrium,
\begin{equation}
	n_0 (p, \mu) = 0.
	\label{eq:dGdmuIsZero}
\end{equation}
\end{subequations}
Indeed, we have $n_0 = - \frac{\partial G_{0}}{\partial \mu} = -
\frac{\mathrm{d} g_0 (p, \lambda_0 (p, \mu), \mu)}{\mathrm{d} \mu} = -
\frac{\partial g_0}{\partial \mu} - \frac{\partial \lambda_0}{\partial
\mu} \frac{\partial g_0}{\partial \lambda}$, where the both terms are
zero by the equilibrium
conditions~(\ref{eq:g0EquiTransv}--\ref{eq:lambda0def}).

To summarize, we view $\lambda$ as an internal variable slaved to the
`macroscopic' variable $\mu$ (the roles of $\lambda$ and $\mu$ could
be exchanged but the other way around would be more complicated as the
mapping from $\lambda$ to $\mu$ is not single-valued).  A catalog of
homogeneous solutions can be obtained by (\emph{i}) solving the axial
equilibrium~(\ref{eq:g0EquiAxial}) for $\lambda = \lambda_0 (p, \mu)$,
(\emph{ii}) defining a reduced potential energy $G_0 (p , \mu)$
by~(\ref{eq:Gdef}), and (\emph{iii}) solving the equilibrium condition
$n_{0}(p, \mu) = 0$ in the $(p,\mu)$ plane.

\begin{figure}
	\centering
	\includegraphics[width=.93\textwidth]{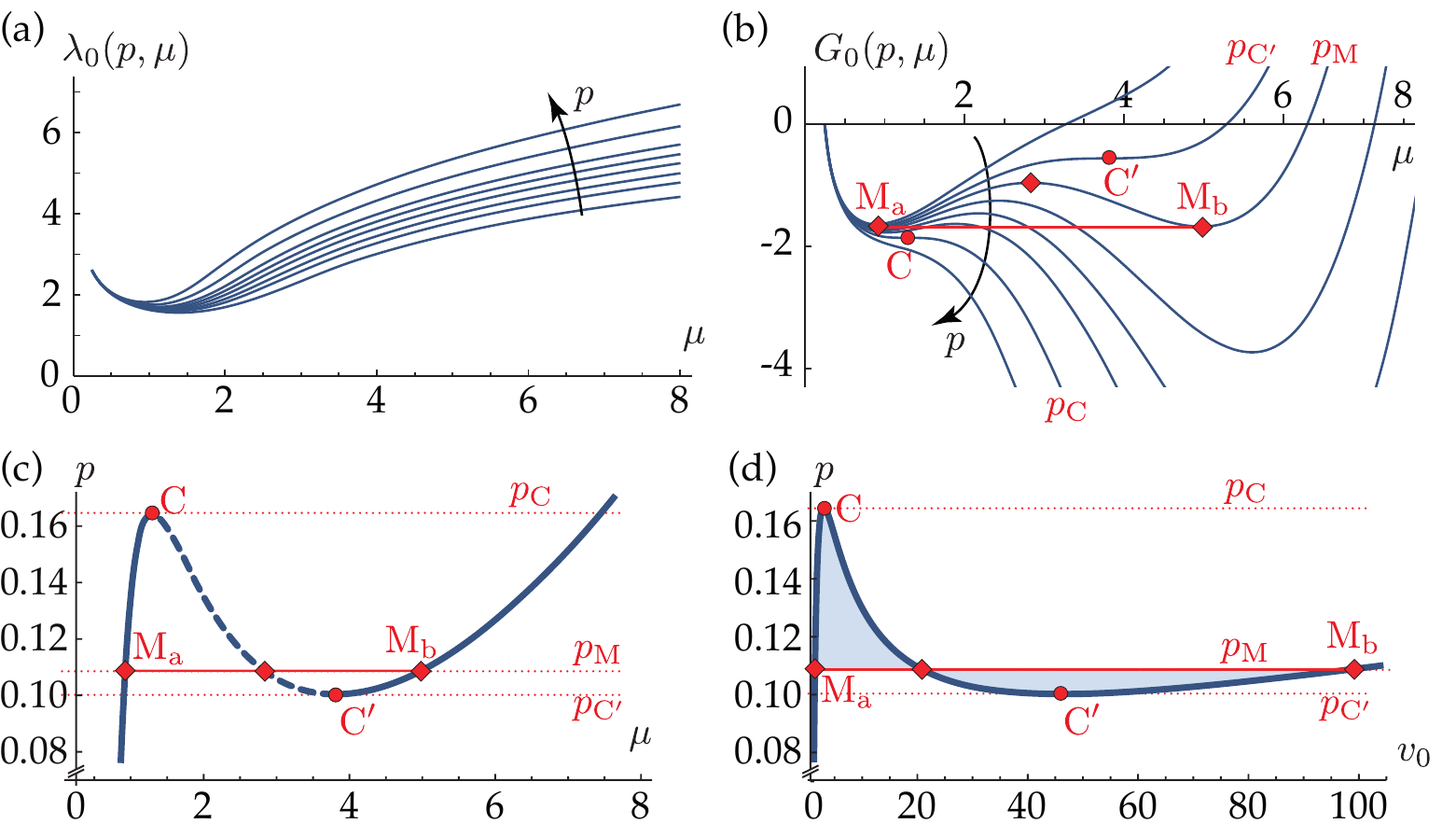}
	\caption{Analysis of the homogeneous solutions, with $F=1.149$.
	(a)~For any value of the pressure $p$, the axial
	equilibrium~(\ref{eq:lambda0def}) yields an implicit curve in the
	$(\mu,\lambda)$ plane that defines the stretch
	$\lambda_{0}(p,\mu)$ in terms of $\mu$; the different curves
	correspond to $p = 0.0885, 0.1002, 0.1087, 0.1187, 0.1285, 0.140,
	0.1646, 0.200$.
	(b)~Reduced potential $G_0$ as a function of $\mu$ for the same
	set of values of $p$.  Critical points are shown in red:
	Consid\`ere points $\mathrm{C}$ and $\mathrm{C}'$ where the
	pressure is extremal (disks), and Maxwell construction (line and
	diamonds).
	%
	%
	(c)~Homogeneous solutions in the $(\mu,p)$-plane, as determined by
	solving the transverse equilibrium $n_{0}(p,\mu)=0$. The dashed 
	part of the curve between the Consid\`ere points is unstable, as 
	$\partial^{2}G_{0}/\partial \mu^{2} < 0$.
	(d)~Same set of homogeneous solutions, now represented in the
	$(v_{0},p)$ plane where $v_{0}=\mu^2 \, \lambda_{0}(p,\mu)$ is the
	ratio of the final to the initial volume.  With this set of
	conjugate variables, Maxwell's rule applies and the two shaded
	regions have the same area.}
	\label{fig:hom_solutions}
\end{figure}

This program has been carried out and the results are shown in
figure~\ref{fig:hom_solutions}.
The homogeneous stretch $\lambda_{0}(p,\mu)$ and the potential
$G_{0}(p,\mu)$ are shown in parts a and b of the figure.  In
figure~\ref{fig:hom_solutions}c, the pressure is plotted in terms of
$\mu$ and is seen to increase, attain a local maximum $p_{\mathrm{C}}=
0.1646$, decrease, attain a local minimum $p_{\mathrm{C}'} = 0.1002$,
and finally increase again.  The points of extremal pressure are where
the onset of localization is expected to occur in a infinite medium
($L=\infty$) according to Consid\`ere's
criterion~\cite{Considere-Memoire-sur-lemploi-du-fer-et-de-lacier-1885}:
we will refer to them as \emph{Consid\`ere points}.  For intermediate
values of the pressure, $p_{\mathrm{C}} < p < p_{\mathrm{C'}}$, the
potential $G_{0}(p,\mu)$ plotted in figure~\ref{fig:hom_solutions}b
has two minima and one maximum as a function of $\mu$.  The
non-convexity of $G_{0}$ makes it possible for the bulged and unbulged
domains to coexist, as recalled in the next section; the diffuse
interface model derived later in~\S\ref{s:reducedmodel} will be able
to account for the boundary between these domains.

\subsection{Maxwell's construction}

In a first attempt to address inhomogeneities, we consider solutions
made up of two phases, with respective properties $\left(
\lambda_{\mathrm{a}} = \lambda_0 \left( p, \mu_{\mathrm{a}} \right), \mu_{
\mathrm{a}} \right)$ and $\left( \lambda_{\mathrm{b}} = \lambda_0 \left(
p, \mu_{\mathrm{b}} \right), \mu_{ \mathrm{b}} \right)$.  Discontinuities
are allowed for the moment, their contribution to the energy being
ignored: gradient term $\mu'$ appearing in the membrane model are
simply discarded.  Let $c$ denote the fraction of the phase
`$\mathrm{a}$', and $(1-c)$ the fraction of the phase `$\mathrm{b}$',
as measured after pulling everything back in the reference configuration.

Under these assumptions, the membrane 
energy~(\ref{eq:completeEnergy}) takes the form
\begin{equation}
	\mathcal{E}_{0} \left( p, c, \mu_{\mathrm{a}}, \mu_{\mathrm{b}}
	\right) = L \left( c\,G_0 \left( p, \mu_{\mathrm{a}} \right) + (1 -
	c) \, G_0 \left( p, \mu_{\mathrm{b}} \right) \right) .
	\nonumber
\end{equation}
Optimizing with respect to the $\mu_i$'s and to $c$ successively, we
find
\begin{equation}
\begin{array}{rclll}
n_0 (p, \mu_i) & = & 0 & \text{\qquad} & \text{(mechanical equilibrium
in each phase)}\\
G_0 \left( p, \mu_{\mathrm{b}} \right) - G_0 \left( p, \mu_{\mathrm{a}}
\right) & = & 0 &  & \text{(chemical equilibrium)}
\end{array} \label{eq:twoPhaseEquilConds}
\end{equation}
These equations can be solved for $p$ and the $\mu_{i}$'s: in
particular this selects a value of the pressure $p = p_{\text{M}}$,
known as Maxwell's pressure, where the two phases can coexist.  The
propagation pressure $p_{\text{M}}$ is a function of both the applied
force $F$ and of the constitutive model for the membrane, but this is
implicit in our notation.  For $F = 1.149$ and for the particular
values of the constitutive parameters used here, we have obtained the
Maxwell load as $p_\mathrm{M} = 0.1087$, see the red line joining the points
labeled $\mathrm{M}_{\mathrm{a}}$ and $\mathrm{M}_{\mathrm{b}}$ in
figure~\ref{fig:hom_solutions}

Maxwell's equal-area rule for the propagation pressure can be
rederived as follows.  The quantity $G_0 ( p, \mu_{\textrm{b}}) - G_0 (
p, \mu_{\textrm{a}})$ appearing in~(\ref{eq:twoPhaseEquilConds}) can be
written as the integral of $\mathrm{d} G_0$ along the curve
corresponding to homogeneous solutions in the $(p,\mu)$ plane.  Along
this curve, $\frac{\partial G_0}{\partial \mu} = - n_0 = 0$
by~(\ref{eq:n0label}).  Therefore, $\mathrm{d} G_0 = \frac{\partial
G_0}{\partial p} (p, \mu)\, \mathrm{d} p = \frac{\mathrm{d} g_0 (p,
\lambda_0 (p, \mu), \mu)}{\mathrm{d} p} \,\mathrm{d} p = \left(
\frac{\partial g_0}{\partial p} + \frac{\partial g_0}{\partial
\lambda} \frac{\partial \lambda_0}{\partial \mu} \right) \mathrm{d} p
= \frac{\partial g_0}{\partial p} \mathrm{d} p $ after
using~(\ref{eq:lambda0def}).  In view of (\ref{eq:H0Decomposition}),
this can be written as $\mathrm{d} G_0 = \frac{e}{2}\,
v_{0}\,\mathrm{d}p$, where $v_{0}(p,\mu)= \lambda_{0}(p,\mu) \mu^2 =
\frac{\pi\,r^{2}\,\mathrm{d}z}{\pi\,R^{2}\,\mathrm{d}Z}$ denotes the
ratio of the deformed to the undeformed volume of homogeneous
solutions.  Using~(\ref{eq:twoPhaseEquilConds}), the variation of
$G_{0}$ from one Maxwell point $\mu_{\mathrm{a}}$ to the other
$\mu_{\mathrm{b}}$ is zero, and so
\begin{equation}
	 \int_{\mu_{\mathrm{a}}}^{\mu_{\mathrm{b}}} v_{0}(p,\mu) \,\mathrm{d} p
	 = 0
	 \textrm{.}
	\nonumber
\end{equation}
This equality implies the equality of the area of the
shaded regions in figure~\ref{fig:hom_solutions}d, which uses $v_{0}$ 
as the horizontal axis and $p$ as the vertical axis.

\section{Derivation of the diffuse interface model}
\label{s:reducedmodel}


We proceed to derive the diffuse interface model from the non-linear
membrane theory.  This reduction combines an assumption of scale
separation, whereby the solution is assumed to vary on a length scale
$L$ much larger than the radius $R$, and the elimination of the
unknown $\lambda$ in favor of $\mu$ by means of the relation $\lambda
= \lambda_{0}(p,\mu)$.

\subsection{Principle of the expansion}

We assume scale separation and use the convention that the radius $R$
is fixed and finite while $L = R/\varepsilon$ goes to infinity: the
solution is sought in terms of a scaled variable $\tilde{Z} =
\varepsilon\,Z$ through scaled functions $\tilde{\lambda}$ and
$\tilde{\mu}$, where $\varepsilon\ll 1$ is our expansion parameter,
\begin{equation}
	\lambda_{\varepsilon}(Z) = \tilde{\lambda}(\varepsilon\,Z),\qquad
	\mu_{\varepsilon}(Z) = \tilde{\mu}(\varepsilon\,Z)
	\textrm{.}
	\nonumber
\end{equation}

As a consequence of this scaling assumption, any derivative with
respect to the slow axial variable $Z$ entails a multiplication by the
small parameter $\varepsilon$:
\begin{alignat}{2}
	&\lambda_{\varepsilon}(Z) = \tilde{\lambda}(\varepsilon\,Z) = 
	\mathcal{O}(1),&\qquad
	&\mu_{\varepsilon}(Z) = \tilde{\mu}(\varepsilon\,Z) = 
	\mathcal{O}(1),\nonumber\\
	&\frac{\mathrm{d}\lambda_{\varepsilon}}{\mathrm{d}Z} = 
	\epsilon\,
	\frac{\mathrm{d}\tilde{\lambda}}{\mathrm{d}\tilde{Z}}(\varepsilon\,Z) = 
	\mathcal{O}(\varepsilon),&\qquad&
	\frac{\mathrm{d}\mu_{\varepsilon}}{\mathrm{d}Z} = 
	\varepsilon\,
	\frac{\mathrm{d}\tilde{\mu}}{\mathrm{d}\tilde{Z}}(\varepsilon\,Z) =
	\mathcal{O}(\varepsilon)
	\textrm{.}
	\nonumber
\end{alignat}
For the sake of legibility, we drop the subscripts $\varepsilon$ and
remove any reference to the scaled functions $\tilde{\lambda}$ and
$\tilde{\mu}$ in the following: it will be sufficient for us to use
the above order of magnitude estimates.

\subsection{Derivation of the gradient effect by a formal expansion}

The general expression of the membrane
strain~(\ref{eq:genericMembraneStrain}) can be split in two terms
\begin{subequations}
	\begin{equation}
		\mathbf{E} (\lambda, \mu, \mu')  =  \mathbf{E}_0 (\lambda, \mu)
		+\mathbf{E}_1 (\mu'),
	\end{equation}
	where the first one depends on the stretch and the second one on the 
	stretch gradient,
	\begin{eqnarray}
		\mathbf{E}_0 (\lambda, \mu) & = & \frac{1}{2}  \left(
		\begin{array}{c}
			\mu^2 - 1\\
			\lambda^2 - 1
		\end{array}\right), \\
		\mathbf{E}_1 (\mu') & = & \frac{1}{2}  \left(
		\begin{array}{c}
			0\\
			(R \mu')^2
		\end{array}\right).  \label{eq:E1}
	\end{eqnarray}
In view of the results from the previous section, their orders of
magnitude are
\begin{equation}
	\mathbf{E}_0 (\lambda, \mu) =\mathcal{O} (1) \qquad \mathbf{E}_1 (\mu')
	=\mathcal{O} (\varepsilon^2).
	\label{eq:mainScalings}
\end{equation}
\end{subequations}
In line with the fact that we use the finite elasticity theory, the
strain $\mathbf{E}_0$ is of order 1.

$\mathbf{E}_1$ being a small correction to $\mathbf{E}_0$, the strain
energy density can be expanded as
\begin{equation}
\begin{array}{rcl}
w (\mathbf{E}) & = & w (\mathbf{E}_0 (\lambda, \mu) +\mathbf{E}_1
(\mu'))\\
& = & w (\mathbf{E}_0 (\lambda, \mu)) +
\frac{\partial w}{\partial \mathbf{E}}(\mathbf{E}_0 (\lambda, \mu))
\cdot \mathbf{E}_1 (\mu') +\mathcal{O}
(|\mathbf{E}_1|^{2})\\
& = & w_0 (\lambda, \mu) +\boldsymbol{\Sigma}_0 (\lambda, \mu) \cdot
\mathbf{E}_1 (\mu') +\mathcal{O} (\varepsilon^4)
\end{array} \label{eq:wOfEExpansion}
\end{equation}
where we have used the definition of the membrane stress
$\boldsymbol{\Sigma}_0$ in~(\ref{eq:sigma0}).  Inserting this
into~(\ref{eq:completeEnergy}) yields the following
approximation of the energy
\begin{equation}
	\mathcal{E}_{\text{memb}} [p, \mu] = \int_{0}^{L} \Big(
	g_0 (p, \lambda (Z), \mu (Z)) 
	+
	\boldsymbol{\Sigma}_0 (\lambda (Z), \mu (Z)) \cdot \mathbf{E}_1 (\mu' (Z))
	\Big)\,
	\mathrm{d} Z 
	+\mathcal{O} (L \, \varepsilon^4). 
	\label{eq:E-tmp1}
\end{equation}
Note that the gradient of axial stretch $\lambda'$ does not appear in
this expression.

\subsection{Energy of the diffuse interface model}

An important result, proved in appendix~\ref{B:Redmodel}, is that it
is consistent, at this order of approximation, to replace the unknown
$\lambda(Z)$ with the axial stretch $\lambda_{0}(p,\mu(Z))$ of the
homogeneous solution having the local value of $\mu(Z)$ as its
circumferential stretch.  This eliminates $\lambda(Z)$ from the
equations, and we obtain the \emph{diffuse interface model}
\begin{subequations}
	\label{subeqs:EfinalAndB0}
\begin{equation}
\mathcal{E} [p, \mu] = \int_{0}^{L} G_0 (p, \mu (Z))\, \mathrm{d} Z + \frac{1}{2}  
\int_{0}^{L} B_{0} (p,
\mu (Z)) \, \mu^{\prime 2} (Z) \, \mathrm{d} Z. 
\label{eq:Efinal}
\end{equation}
where we have omitted terms of order $L\,\varepsilon^{4}$ and higher.
The coefficient $B_{0}$ of the regularizing term has a simple
expression which is found by identifying with~(\ref{eq:E-tmp1})
and~(\ref{eq:sigma0}) as
\begin{equation}
B_{0} (p, \mu(Z)) = R^2  \, \left[ \frac{1}{\lambda}  \, \frac{\partial
	w_{0}}{\partial \lambda} (\lambda, \mu(Z)) \right]_{\lambda = \lambda_0 (p,
	\mu(Z))}.
 \label{eq:Bexpression}
\end{equation}
\end{subequations}
This defines the regularizing term in terms of the energy
$w_{0}(\lambda,\mu)$ of homogeneous solutions,
see~(\ref{eq:strainEnergy}).  Even though this is implicit in our
notation, both $G_{0}$ and $B_{0}$ depend on the force $F$.

Equations~(\ref{eq:Efinal}--\ref{eq:Bexpression}) are our main result,
and can be restated as follows.  The energy $\mathcal{E}_{\text{memb}}$ of the full
non-linear membrane model can be approximated as the sum of
({\emph{i}})~the non-regularized energy $\int G_{0}\,\mathrm{d}Z$
which depends on the stretch $\mu$ but not on its gradient, and is of
order $L$, and ({\emph{ii}})~a much smaller correction $\frac{1}{2}
\int B_0 \, \mu^{\prime 2} \mathrm{d} Z$, of order $\varepsilon^{2}$, that
depends on the strain $\mu$ and as well as on its gradient $\mu' = \frac{\mathrm{d} \mu}{\mathrm{d} Z}$.  These two terms provide an approximation of the full energy
$\mathcal{E}$ of the non-linear membrane model which is accurate up to
order $L\,\varepsilon^4$.

\subsection{Non-linear equilibrium of the diffuse interface model}

The equilibrium equations are obtained from~(\ref{eq:Efinal}) by the
Euler-Lagrange method as
\begin{subequations}
	\label{subeqs:BVpbDiffuseInterfacee}
\begin{equation}
n_0 (p, \mu (Z)) - \frac{1}{2} \, \frac{\partial B_{0}}{\partial \mu} (p, \mu
(Z)) \,  \mu^{\prime 2} (Z) + \frac{\mathrm{d}}{\mathrm{d} Z} (B_{0} (p, \mu (Z))
\, \mu' (Z))
= 0. \label{eq:equil2dGrdModel}
\end{equation}
In the absence of kinematic constraints, the variational method yields
the natural conditions at the endpoints as well,
\begin{equation}
	\mu'(0) = \mu'(L) = 0
	\textrm{.}
	\label{eq:BC2dGrdModel}
\end{equation}
\end{subequations}
Here, $\mu'(L) = 0$ is consistent with the symmetry condition at the
center $Z=0$ of the bulge.

The equilibrium condition~(\ref{eq:equil2dGrdModel}) reduces to the
condition~(\ref{eq:dGdmuIsZero}) applicable to homogeneous solutions,
namely $n_0 (p, \mu) = 0$, when the gradient effect is removed, by
setting $B_{0} = 0$.

\subsection{Solution for a domain boundary in an infinite balloon}

The existence of a first integral associated with the
equilibrium~(\ref{eq:equil2dGrdModel}) has been noted by a number of
authors such
as Coleman \& Newman~\cite{ColemanNewman}.  It
can be obtained by expanding the derivative in the last term in the
right-hand side, and by multiplying the entire side by $\mu'(Z)$; the
result is $\frac{\mathrm{d}(-G_{0}+B_{0}\,\mu'^{2})}{\mathrm{d}Z} =
0$.  This shows that the following quantity is conserved:
\begin{equation}
	-G_{0}(p,\mu(Z))+B_{0}(p,\mu(Z))\,\mu'^{2}(Z) = C
	\textrm{.}
	\label{eq:conservedIntegral}
\end{equation}
This equation can be used to solve for $\mu(Z)$ by quadrature.
However, this method is impractical for numerical calculations
as it involves evaluating integrals that are close to singular,
even when the singular parts are taken care of analytically~\cite{%
BAJH}.  This is why our numerical simulations
in~\S\ref{pre-ssec:fullSimulations}\ref{ssec:fullSimulations} use a
direct integration method of the
equilibrium~(\ref{eq:equil2dGrdModel}) rather than the quadrature
method.

In the case of the boundary separating two domains in an infinite
medium, however, the quadrature method is tractable.  Then, the
pressure matches Maxwell's pressure, $p=p_{\mathrm{M}}$, and $\mu(Z)$
tends to $\mu_{\mathrm{a}}$ and $\mu_{\mathrm{b}}$ for $Z\to
\pm\infty$, respectively.  The value of $C$ consistent with these
asymptotic behaviors is the common value $C =
G_{0}(p_{\mathrm{M}},\mu_{\mathrm{a}}) =
G_{0}(p_{\mathrm{M}},\mu_{\mathrm{b}})$ of the potential,
see~(\ref{eq:twoPhaseEquilConds}).  The implicit
equation~(\ref{eq:conservedIntegral}) can then be plotted in the phase
space $(\mu(Z),\mu'(Z))$ using a contour plot method.  We have checked
that the resulting curve (not shown) falls on top of the dotted green
curve labeled $\mathrm{A}'$ in
figure~\ref{fig:fullnonlinear_solutions}c, obtained by numerical
integration of the equilibrium with a large but finite aspect ratio,
$L/R=30$: the analytical solution~(\ref{eq:conservedIntegral}) in an
infinite balloon provides an excellent approximation to a propagating
interface in a finite balloon, as long as it is sufficiently remote
from the endpoints.  In the bifurcation diagram, the numerical
solutions appears as a point $\mathrm{A}'$ lying almost exactly on
Maxwell's plateau, see figure~\ref{fig:fullnonlinear_solutions}a.

This analytical solution is also an excellent approximation to the
domain boundary predicted by the original membrane model, as discussed
below in~\S\ref{pre-ssec:fullSimulations}\ref{ssec:fullSimulations}.

\section{Comparison of the diffuse interface and membrane models}
\label{sec:application}
\label{pre-ssec:fullSimulations}

Using a formal expansion method, we have shown that the 2d non-linear
axisymmetric membrane model (\S\ref{s:nonlinearMembraneModel}) is
asymptotically equivalent to the 1d diffuse interface model
in~(\ref{subeqs:EfinalAndB0}).  This equivalence holds for `slowly'
varying solutions, \emph{i.e.}\ when the axial gradients involve a
length scale much larger than the tube radius,
$|\mathrm{d}\mu/\mathrm{d}Z|\ll 1/R$.  Here, we compare the
predictions of the approximate diffuse interface model to those of the
original membrane model.  The goal is twofold.  First, we verify our
asymptotic expansion by checking consistency for slowly-varying
solutions.  Second, we push the diffuse interface model outside its
domain of strict mathematical validity, by applying it to problems
involving sharp boundaries and
comparing to the predictions of the original membrane model. 

\subsection{Comparison of the full bifurcation diagrams}
\label{ssec:fullSimulations}

\begin{figure}
	\centering
	\includegraphics[scale=.85]{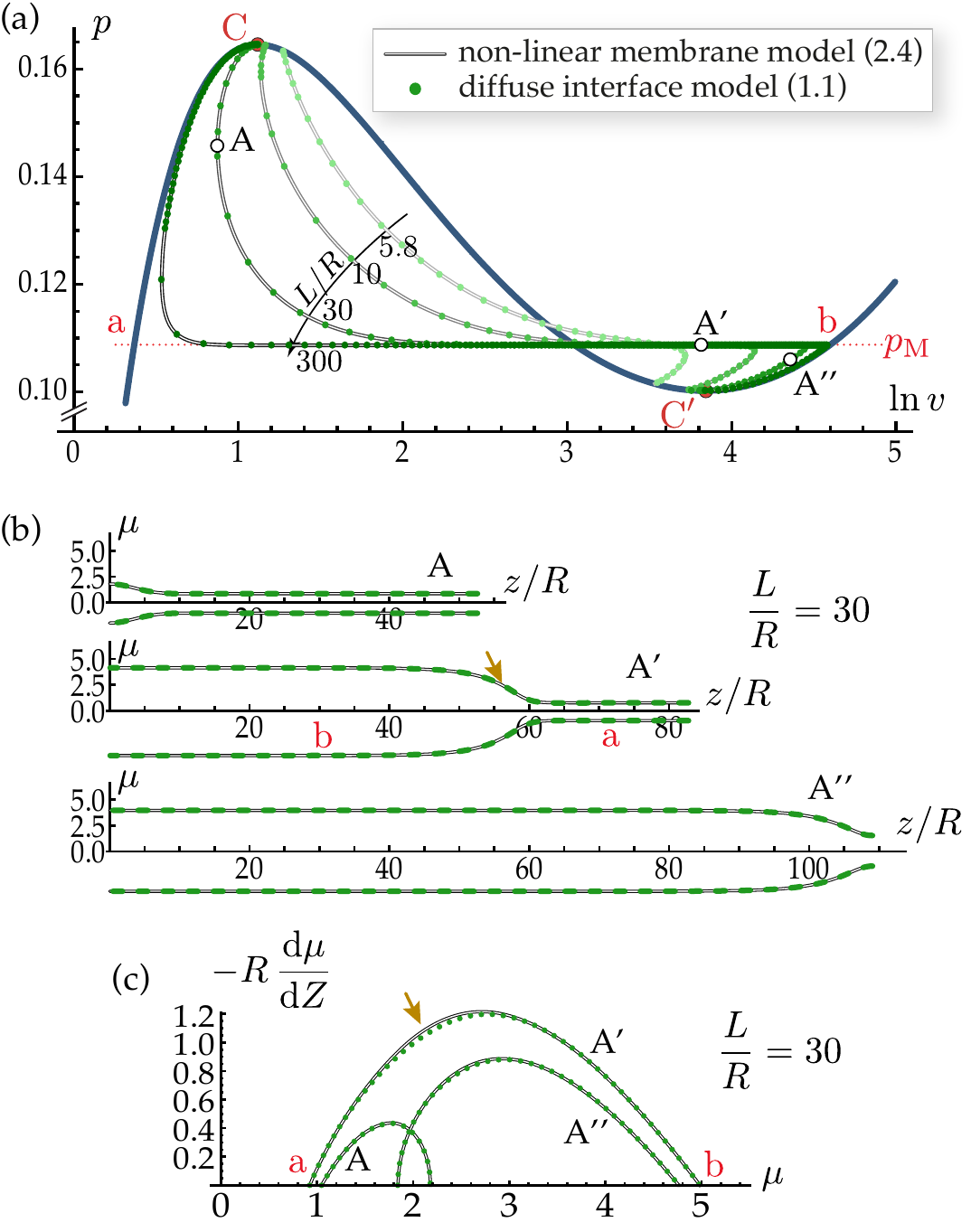}
	\caption{Comparison of the predictions of the non-linear membrane
	model and of the diffuse interface model, for $F=1.149$.  (a)
	Bifurcation diagrams for different values of the aspect-ratio
	$L/R$: homogeneous solutions (dark blue curve), bifurcated
	branches of the membrane model (double-struck black curves) and of the
	diffuse interface model (green dots); Consid\`ere's points (red
	dots) and Maxwell's plateau (dotted red line).  Note that we use
	the \emph{logarithm} of the scaled volume $v$ on the horizontal
	axis, so that Maxwell's equal-area rule does not apply directly.
	(b) Comparison of the deformed configurations in physical space
	$(z/R,\mu=r/R)$ for $L/R = 30$, for the three configurations
	labeled $\mathrm{A}$, $\mathrm{A}'$ and $\mathrm{A}''$ in (a),
	corresponding to $(v,p) = (2.39, 0.146)$, $(45., 0.109)$ and
	$(77.43, 0.106)$ respectively.  (c) Same solutions visualized in
	the phase space: a small discrepancy is visible in the center of
	the sharp interface (arrow).}
	\label{fig:fullnonlinear_solutions}
\end{figure}

We start by comparing the bifurcation diagrams obtained with each one
of the models for balloons of finite length $L$, see
figure~\ref{fig:fullnonlinear_solutions}a.
In this numerical example, both the membrane and the diffuse interval
models use the constitutive law in~(\ref{eq:ogdenlaw}), the standard
set of material parameters listed below this equation, and the value
of the pulling force $F$ in~(\ref{eq:forceF}).  We limit our attention
to solutions that are either homogeneous or comprise a single bulge
centered at $Z=0$: recall that the simulation domain $(0,L)$
represents one half of a real balloon.

The equilibrium equations of the membrane model are obtained from the
energy~(\ref{eq:completeEnergy}) by an Euler-Lagrange method, and are
solved numerically.  These equations of equilibrium and their
numerical solution have already been documented
in~\cite{Kyriakides-Chang-The-initiation-and-propagation-of-a-localized-1991},
and we refer to this work for details; in our work, the solution
branches were calculated using the path-following
method from the AUTO-07p library~%
\cite{Doedel-Champneys-EtAl-AUTO-07p:-continuation-and-bifurcation-2007}.
While
\cite{Kyriakides-Chang-The-initiation-and-propagation-of-a-localized-1991}
used boundary conditions representing rigid plugs, we use instead the
natural boundary conditions, 
namely the axial and radial equilibria $F=\frac{\partial w}{\partial
\ell_z} \, \frac{z'}{\ell_z} - p \, e \frac{\ell_\theta^2}{2}$ and
$\mu'=0$.  These boundary conditions are relevant to the soft boundary
device sketched in figure~\ref{fig:experiment}b, and are enforced at
$Z=L$.  At the center of bulge $Z=0$, we impose the symmetry
conditions $\mu'=0$ and $z=0$.

To solve the diffuse interface
model~(\ref{subeqs:BVpbDiffuseInterfacee}) numerically, we first
sample the functions $n_{0}(p,\mu)$ and $B_{0}(p,\mu)$ numerically.
This tabulation is available as by-product of the analysis of
homogeneous solutions from Section~\ref{s:homsol_barmod}.  Next, the
solution branches are generated by solving the boundary value
problem~(\ref{subeqs:BVpbDiffuseInterfacee}) using the path-following
library AUTO-07p.  Alternatively, we tried solving this boundary value
problem by the quadrature method described earlier, but it did not work
well for the reason already explained.


The bifurcation diagrams are shown in
figure~\ref{fig:fullnonlinear_solutions}a.  The homogeneous solutions
are plotted using the thick, dark blue curve: they are identical for
both models, and are also identical to those derived earlier in
figure~\ref{fig:hom_solutions}d.  Bifurcated solutions are shown as
black double-struck curves (membrane model) and green dots (diffuse
interface model) for different value of $\overline{L} = L/R$.  The
bifurcation diagram uses the natural logarithm of the scaled volume
$v$ on the horizontal axis,
\begin{equation}
	v = \frac{1}{L/R}\,\int_0^{L} \mu^2 \,
	\lambda_{0}(p,\mu)\,\frac{\mathrm{d}Z}{R}
	\textrm{.}
	\label{eq:v}
\end{equation}
This is consistent with the definition of the scaled volume $v_{0}$
used in the analysis of homogeneous solutions.  For large values of the
aspect-ratio $L/R$, the bifurcated branches display a plateau 
corresponding to Maxwell's pressure $p_{\mathrm{M}}$.

The diffuse interface model appears to be highly accurate, as its
bifurcation diagram is almost identical to that of the membrane model:
in the figure, the green dots fall exactly onto the double-struck
curves.  Given that the diffuse interface model has been derived under
an assumption of `slow' axial variations, it could be expected that
the models would agree near the bifurcation points (in the
neighborhood of the dark blue curve) where the localization is mild.
We did not anticipate the good agreement far from the bifurcation
point, for configurations featuring relatively sharp interfaces such
as that labeled $\mathrm{A}'$ in the figure: for this solution, the
largest value of the stretch gradient is 1.2, see
figure~\ref{fig:fullnonlinear_solutions}c---even though this is not a
small number, the diffuse interface model remains remarkably accurate.

Selected deformed configurations are plotted in
figure~\ref{fig:fullnonlinear_solutions}b in real space: the
predictions of both models are still indistinguishable, even inside
the domain boundary.  The predictions of the two models are not
exactly identical, however: a small difference is visible when these
solutions are represented in phase space, see
figure~\ref{fig:fullnonlinear_solutions}c; in phase space, the subtle
features of the interface are highlighted, while the uniform domains
shrink to the points labeled `$\mathrm{a}$' and `$\mathrm{b}$' in the
figure.

To sum up, the diffuse interface model reproduces the entire
bifurcation diagram of the original membrane model with good accuracy,
even for well localized domain boundaries.  In the following sections,
we show that it is also well suited to linear and non-linear buckling
analyses.

\subsection{Onset of bulging: linear bifurcation analysis, finite length}

We now compare the bifurcation load at the onset of bulging, as
predicted by the diffuse interface model on the one hand, and by the
membrane model on the other hand.  The diffuse interface model yields
a simple analytical prediction, that matches that of the membrane
model exactly.

\begin{figure}
	\centering
	\includegraphics[width=.99\textwidth]{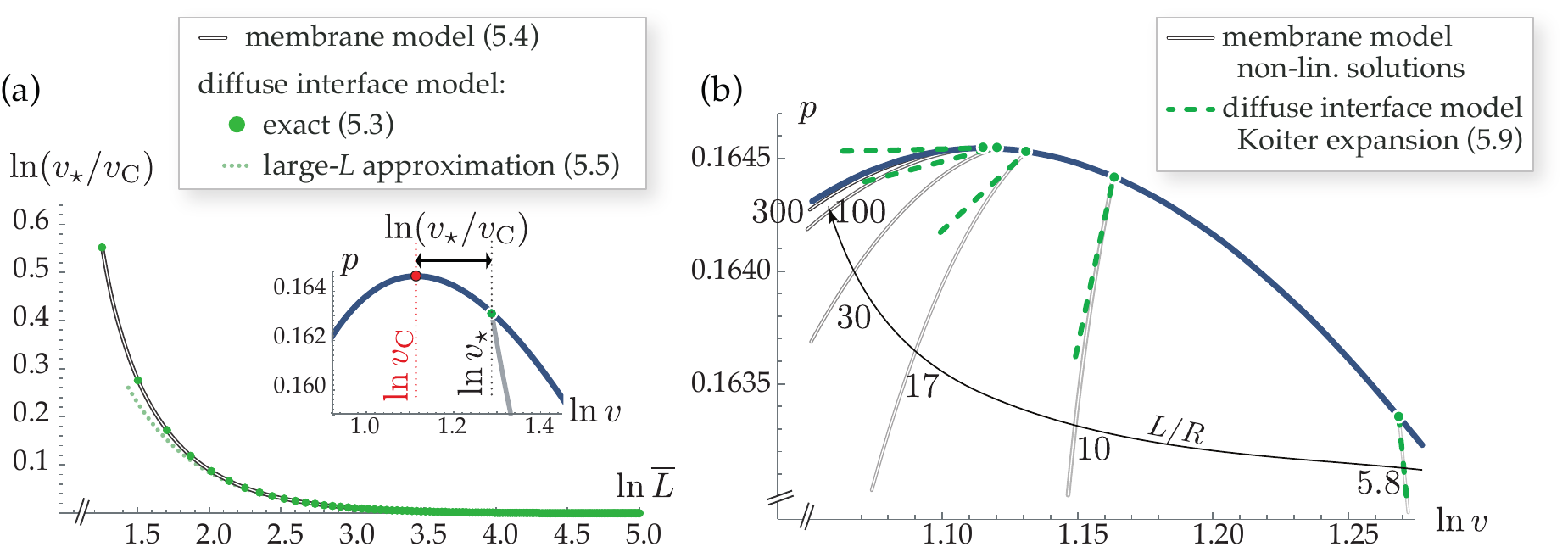}
	\caption{Linear and non-linear bifurcation analyses of the bulging instability
	based on the diffuse interface model, and comparison with the
	membrane model.  (a) Critical volume $v_\star$ at the onset of
	bulging as a function of the aspect-ratio $\overline{L} = L/R$.
	The diffuse interface model predicts the bifurcation load exactly
	(the green dots fall exactly onto the double-struck curve).
	\emph{Inset}: bifurcation point shown in the bifurcation diagram
	as in figure~\ref{fig:fullnonlinear_solutions}a for a particular
	value $\overline{L} = L/R = 5.8$.  (b) Initial tangents to the
	bifurcated branches for different values of the aspect-ratio
	$\overline{L} = L/R$, as predicted by a weakly non-linear
	expansion of the membrane model (green dotted lines); comparison
	with the non-linear branches predicted by the membrane model
	(double-struck curves).}
	\label{fig:onset_bifurc}
\end{figure}

The critical load of the diffuse interface model is derived by a
classical linear bifurcation analysis as follows.  Consider a
perturbation to a homogeneous solution $\mu_{0}$, in the form $\mu(Z)
= \mu_{0} + \mu_{1}(Z)$.  Linearizing the equilibrium equation of the
diffuse interface model in~(\ref{eq:equil2dGrdModel}) with respect to
$\mu_{1}$, we obtain
\begin{equation}
	\frac{\partial n_0}{\partial \mu} (p, \mu_{0}) \, \mu_1(Z)  + B_{0} (p, 
	\mu_{0})\,
	\mu_1'' (Z) = 0.
	\label{eq:bifurcation_cond}
\end{equation}
The boundary conditions are $\mu_1'(0)=0$ and $\mu_1'(L)=0$.  The
first critical mode $\mu_1(Z) = \cos \frac{\pi\,Z}{L}$ corresponds to
half a bulge in the simulation domain $(0, L)$.  When inserted into
the above expression, this yields
\begin{equation}
	\frac{\partial n_0}{\partial \mu} (p, \mu_{0})=\frac{B_{0}(p, \mu_{0})}{R^2}  \, 
	\frac{\pi^2}{\overline{L}^2}
	\quad
	\textrm{(diffuse interface model)}
	\textrm{.}
	\label{eq:stability_secondgrad}
\end{equation}
This equation must be solved together with the axial equilibrium 
condition for the unperturbed solution~(\ref{eq:dGdmuIsZero}),
$n_{0}(p,\mu_{0})=0$.  For any given value of the aspect-ratio
$\overline{L}$, the roots $(p_{\star}(\overline{L}),
\mu_{\star}(\overline{L}) = \mu_{0}(p_{\star}(\overline{L})))$ of these two
equations define the critical parameters where the bifurcation occurs.
The corresponding scaled volume can then be reconstructed as
$v_{\star}(\overline{L}) = v_{0}(p_{\mathrm{\star}}(\overline{L}),
\mu_{\star}(\overline{L}))$.  The dependence of the critical volume
$v_{\star}$ on the aspect-ratio is shown by the green dots in
figure~\ref{fig:onset_bifurc}a.

For comparison, we derive the bifurcation load predicted by the
membrane model.  The critical load for hard plugs has been obtained
by Kyriakides \& Chang~\cite{Kyriakides-Chang-The-initiation-and-propagation-of-a-localized-1991}.
Adapting their bifurcation analysis to soft plugs, we obtain the
bifurcation condition as
\begin{equation}
	\lambda_0(p,\mu_0)\,
	\frac{
	(p \, e \, \lambda_0(p, \mu_0)
	-
	w_{0,\mu^{2}}
	)\,
	w_{0,\lambda^{2}}
	+ 
	(
	w_{0,\lambda\mu}
	-
	p \, e \, \mu_0
	)^2
	}{
	w_{0,\lambda}
	\,
	w_{0,\lambda^{2}}
	}
	=
	\frac{\pi^2}{\overline{L}^2}
	\quad
	\textrm{(membrane model)}
	\label{eq:bifurcationMembraneModel}
\end{equation}
where commas in subscripts denote partial derivatives of the
homogeneous strain energy $w_{0}$ defined in~(\ref{eq:strainEnergy}),
all of which are evaluated at $(\lambda,\mu) =
(\lambda_{0}(p,\mu_{0}),\mu_{0})$.  Solving this equation together
with the axial equilibrium yields $p_{\star}$ and $\mu_{\star} =
\mu_{0}(p_{\star})$ as a function of $\overline{L}$, as earlier.
%

The bifurcation loads of the diffuse interface model (green dots) and
of the membrane model (solid dark gray curve) are compared in
figure~\ref{fig:onset_bifurc}a.  They agree exactly.  This is not
surprising as, close to bifurcation, the solutions of the membrane
model depart from a uniform solution by an arbitrarily small
perturbation, implying that the axial gradients are arbitrarily small:
the assumptions underlying the diffuse interface model are satisfied
close to the bifurcation point.  The diffuse interface model captures
exactly the retardation of the onset of buckling in balloons of finite
length.  Similarly, the critical load predicted by the one-dimensional
diffuse interface model for the analysis of necking in solid cylinders
has been found in~\cite{BAJH} to agree exactly with that based on the
three-dimensional analysis
of~\cite{Hutchinson-Miles-Bifurcation-analysis-of-the-onset-1974}.

For large values of the aspect-ratios $\overline{L}$, the bifurcation
equation~(\ref{eq:stability_secondgrad}) can be simplified by noticing
that the left-hand side goes to zero, as the bifurcations takes place
closer and closer to the Consid\`ere point $(p_{\mathrm{C}},
\mu_{\mathrm{C}})$ where the load is maximum, $\frac{\partial
n_0}{\partial \mu} (p_{\text{C}}, \mu_{\text{C}})=0$.  Expanding the
left-hand side accordingly, one obtains
from~(\ref{eq:stability_secondgrad})
\begin{equation}
	\mu_\star \approx \mu_{\text{C}} +  \frac{B_{0} (p_{\text{C}},
	\mu_{\text{C}})}{R^2\,
		\frac{\partial^2
	n_0}{\partial \mu^2} (p_{\text{C}}, \mu_{\text{C}})
	}\, \frac{\pi^2}{\overline{L}^2} 
	\quad\textrm{(diffuse interface model, large $\overline{L}$).}
	\label{eq:stability_secondgrad_considere}
\end{equation}
This yields the dotted curve in figure~\ref{fig:onset_bifurc}a,
which is indeed consistent with the two other curves in the limit
$\overline{L}\to\infty$.

\subsection{Onset of bulging: weakly non-linear bifurcation analysis, finite length}

Following the method of Lyapunov and
Koiter~\cite{Koiter-On-the-stability-of-elastic-equilibrium-1965}, an
expansion of the bifurcated branch can be found by introducing a small
arc-length parameter $\eta$ and expanding $p$ and $\mu$ as
\begin{subequations}
	\label{eq:weakNLexp}
	\begin{alignat}{4}
	\mu &= \mu_0(p) &\;+\;& \eta \, \mu_1 (Z) &\;+\;& \eta^2 \, \mu_2(Z) 
	&\;+\;&\mathcal{O}(\eta^3)\\
	p &=  p_{ \star} && &\;+\;&  \eta^2 \, p_2 &\;+\;&\mathcal{O}(\eta^3),
	\label{eq:pExpansionInTermsOfEta}
	\end{alignat}
\end{subequations}
where $\mu_{0}(p)$ denotes the branch of homogeneous solutions
satisfying the equilibrium condition $n_{0}(p,\mu_{0}(p))=0$,
see~(\ref{eq:dGdmuIsZero}), $p_{\star} = p_{\star} (\overline{L})$ is
the critical pressure found by the linear stability analysis,
$\mu_1(Z) = \cos \frac{\pi Z}{L}$ is the linear bifurcation mode, and
$\mu_{2}(Z)$ and $p_{2}$ are higher-order corrections.  The latter are
now determined by inserting this expansion into the non-linear
equilibrium~(\ref{subeqs:BVpbDiffuseInterfacee}), and by solving it
order by order in $\eta$.

It is actually preferable to work with the weak form of the equilibrium
(principle of virtual work), which formally writes
$\mathcal{E}_{,\mu}(p, \mu )\cdot[\hat{\mu}]=0$, for any kinematically
admissible virtual stretch $\hat{\mu}$.  Here,  
$\mathcal{E}_{,\mu}(p, \mu )$ denotes the first variation of the 
total potential energy, defined as
\begin{equation}
	\mathcal{E}_{,\mu}(p, \mu )\cdot[\hat{\mu}]=
\lim_{t \rightarrow 0} \frac{\mathcal{E} (p,\mu + t\,
	\widehat{\mu}) -\mathcal{E} (p,\mu)}{t} 
	\textrm{.}
	\nonumber
\end{equation}
Higher-order variations of the energy are defined similarly.

When the expansion~(\ref{eq:weakNLexp}) is inserted into the 
principle of virtual work, one obtains, at order $\eta$, the 
condition 
\begin{equation}
	\forall \hat{\mu},\quad
	\mathcal{E}_{,\mu^{2}}(p_\star, 
	\mu_0(p_\star))\cdot[\mu_1,\hat{\mu}]=0.
	\label{eq:mu1}
\end{equation}
We have recovered the bifurcation
condition~(\ref{eq:bifurcation_cond}), which is automatically
satisfied by the linear mode $\mu_1(Z)$.  

At order $\eta^2$ the expansion yields
\begin{equation}
	\forall\hat{\mu},\quad
	\mathcal{E}_{,\mu^{2}}(p_\star, \mu_0(p_\star) )\cdot[\mu_2,\hat{\mu}] 
	+
	\frac{1}{2}\mathcal{E}_{,\mu^{3}}(p_\star, \mu_0(p_\star) )\cdot[\mu_1,\mu_1,\hat{\mu}]=0.
\label{eq:mu2}
\end{equation}
The first term in the left-hand side involves the tangent stiffness
operator $\mathcal{E}_{,\mu^{2}}(p_\star, \mu_0(p_\star) )$ which is
known to be singular by the bifurcation condition~(\ref{eq:mu1}).
Therefore a solvability condition must be verified before attempting
to solve equation~(\ref{eq:mu2}) for $\mu_2(Z)$; it is derived by
replacing $\hat{\mu}$ with $\mu_{1}$ and reads
$\mathcal{E}_{,\mu^{3}}(p_\star, \mu_{0}(p_\star)
)\cdot[\mu_1,\mu_1,\mu_1]=0$.  In the left-hand side, the interactions
of the three modes $\mu_{1}$ produces harmonic waves with wave-vector
$\pi/L$ and $3\pi/L$, which all cancel out upon integration over the
domain $0\leq Z\leq L$: the solvability condition is automatically
verified (this is referred to as a symmetric system in bifurcation
theory).

Next, equation~(\ref{eq:mu2}) can be solved for $\mu_{2}$, and the 
solution reads
\begin{equation}
	\mu_2(Z) = \mu_{20}
	+\mu_{21} \cos (\frac{\pi}{L}\, Z)
	+\mu_{22} \cos (\frac{2\,\pi}{L}\, Z)
	\textrm{.}
	\nonumber
\end{equation}
The coefficient $\mu_{21}$ remains unspecified at this order (in can
be used to re-normalize the arc-length $\eta$), and the other
coefficients are found as
\begin{equation}
	\mu_{20} = 
	\frac{1}{4}\,
	\left(
	\frac{B_{0,\mu}^\star}{B_{0}^\star}
	-
	\frac{n_{0,\mu^{2}}^\star}{n_{0,\mu}^\star}
	\right),\qquad
	\mu_{22} = 
	\frac{1}{12}\, 
	\left(
	-\frac{3\,B_{0,\mu}^\star}{B^\star_{0}}
	+
	\frac{n_{0,\mu^{2}}^\star}{n_{0,\mu}^\star}
	\right),
	\nonumber
\end{equation}
In our notation, any quantity bearing a star is evaluated at the
critical point $(p_\star,\mu_0(p_\star))$.

Next, the solvability condition at order $\eta^3$ yields the
coefficient $p_2$ as
\begin{equation}
	p_2 = 
	\frac{-3 (\lambda_{0,\mu}^\star \, n_{0,\mu^{2}}^\star)^2
	-6\,
	\lambda_0^\star \,
	\lambda_{0,\mu}^\star\,
	n_{0,\mu}^\star\,
	n_{0,\mu^{2}}^\star
	+\lambda_{0}^\star\,
	(
	6\,
	\lambda_{0,\mu^{2}}^\star \,
	(n_{0,\mu})^2
	-3\, \lambda_{0}^\star\,
	n_{0,\mu^{3}}^\star\,
	n_{0,\mu}^\star
	+
	5\,
	\lambda_0^\star\,
	(n_{0,\mu^{2}}^\star)^2
	)
	}{
	24\, 
	\lambda_{0}^\star\,
	\left(
	-\lambda_{0,p}^\star\, (n_{0,\mu}^\star)^2+n_{0,\mu}^\star\,
	\left(\lambda_{0,\mu}^\star\, n_{0,\mu}^\star
	+\lambda_0^\star\,
	n_{0,\mu\,p}^\star
	\right)
	-\lambda_0^\star\, n_{0,\mu^{2}}^\star\,
	n_{0,p}^\star
	\right)
	}
	\textrm{.}  
	\nonumber
\end{equation}
The right-hand side is defined in terms of the properties of the
homogeneous solution (\S\ref{s:homsol_barmod}), and can evaluated
numerically for any value of $\overline{L}$: recall that all the 
quantities in the right-hand side are evaluated at 
$(p_{\star}(\overline{L}), \mu_{\star}(\overline{L}) = \mu_{0}(
p_{\star}(\overline{L}))$, where $p_{\star}(\overline{L})$ is the 
critical load as determined from the linear buckling analysis, 
see~(\ref{eq:stability_secondgrad}).

Finally, an expansion of the volume factor $v$ defined in~(\ref{eq:v})
is obtained as follows.  Observe that the integrand defining $v$ is
the function $v_{0}(p,\mu) = \mu^{2}\,\lambda_{0}(p,\mu)$: inserting
the expansions of $p$ and $\mu$ from~(\ref{eq:weakNLexp}) into $v_{0}$
and averaging over $Z$, one derives an expansion of the volume factor
as $v(p,\mu) = v(p_{\star}, \mu_{\star}) + v_{2}\,\eta^{2} + \dots$,
where the coefficient reads
\begin{equation}
	v_{2} = 
	p_{2}\,v_{0,p}^{\star}
	+\left(\mu_{20}+p_{2}\,\frac{\mathrm{d}\mu_0}{\mathrm{d}p}(p_{\star})\right)
	\,v_{0,\mu}^{\star}
	+\frac{1}{4}\,v_{0,\mu^{2}}^{\star}.
	\nonumber
\end{equation}
The right-hand side depends on the properties of the homogeneous
solutions, and can be evaluated numerically for any given value of the
aspect-ratio $\overline{L}$.

When the expansion of $v$ is combined
with that of $p$ in~(\ref{eq:pExpansionInTermsOfEta}), we finally
obtain the initial slope of the bifurcated branch as
\begin{equation}
	\left(
	\frac{\mathrm{d}p}{\mathrm{d}v}
	\right)_{\star}
	=
	\frac{p_{2}\,\eta^{2}+\cdots}{v_{2}\,\eta^{2}+\cdots}
	=
	\frac{p_{2}}{v_{2}}
	\textrm{,}
\end{equation}
where $p_{2}$ and $v_{2}$ have just been calculated.  The
corresponding tangents are plotted on figure~\ref{fig:onset_bifurc}(b)
for various values of $\overline{L}$.  They agree very well with the
bifurcated branches of the non-linear membrane model.  

To sum up, the
diffuse interface model is amenable to a weakly non-linear analysis
which reproduces accurately the solutions of the original, fully
non-linear membrane model.

\subsection{Onset of localization: weakly non-linear analysis, infinite length}

The domain of validity of the weakly non-linear expansion derived in
the previous section is more and more limited when the aspect-ratio
gets larger, $L/R\to\infty$: in figure~\ref{fig:onset_bifurc}b, the
domain where the tangent (dashed green line) yields a reasonable
approximation to the bifurcated branch (black double-struck curve)
shrinks when the aspect-ratio increases from $L/R=10$ to $300$.  This
is because for large values of $L/R$, the extended buckling mode
localizes rapidly after bifurcation, a feature not captured by the
analysis of the previous section.  Here, we derive a different weakly
non-linear solution of the diffuse interface model, assuming that the
cylinder is infinitely long, $L/R=\infty$.  This solution captures the
quick localization of the bulges; it is similar to that derived Fu
\textit{et
al.}~\cite{Fu-Pearce-EtAl-Post-bifurcation-analysis-of-a-thin-walled-2008}
based on the full membrane model but its derivation is somewhat
simpler.

In the limit $L/R\to \infty$, the bifurcation takes place at the
Consid\`ere point $(\mu_{\mathrm{C}},p_{\mathrm{C}})$, where the
pressure attains its maximum (the other bifurcation taking
place at the minimum pressure $p_{\mathrm{C'}}$, which can be treated
similarly).  Accordingly, the weakly bulged solution satisfies
$p\approx p_{\mathrm{C}}$ and $\mu(Z)\approx \mu_{\mathrm{C}}$, and
the potential $G_{0}$ can be expanded as
\begin{multline}
	G_{0}(p,\mu) = G_{0}^{\mathrm{C}} + 
	G_{0,p}^\mathrm{C}\,(p-p_{\mathrm{C}}) + 
	G_{0,\mu}^\mathrm{C}\,(\mu-\mu_{\mathrm{C}}) \cdots \\
	{} + 
	\frac{G_{0,p^{2}}^\mathrm{C}}{2}\,(p-p_{\mathrm{C}})^{2} +
	G_{0,p\mu}^\mathrm{C}\,(p-p_{\mathrm{C}})\,(\mu-\mu_{\mathrm{C}}) + 
	\frac{G_{0,\mu^{2}}^\mathrm{C}}{2}\,(\mu-\mu_{\mathrm{C}})^{2} +
	\frac{G_{0,\mu^{3}}^\mathrm{C}}{6}\,(\mu-\mu_{\mathrm{C}})^{3} +
	\cdots
	\nonumber
\end{multline}
The arguments appearing in subscript after a comma denote a partial
derivative, while a superscript `$\mathrm{C}$' means that the function
is evaluated at Consid\`ere's point
$(p_{\mathrm{C}},\mu_{\mathrm{C}})$.  The values of all the
coefficients $G_{0}^{\mathrm{C}}$, $G_{0,p}^\mathrm{C}$, etc.\ are available
from the analysis of homogeneous solutions (\S\ref{s:homsol_barmod}).

\begin{figure}
	\centerline{\includegraphics[scale=.87]{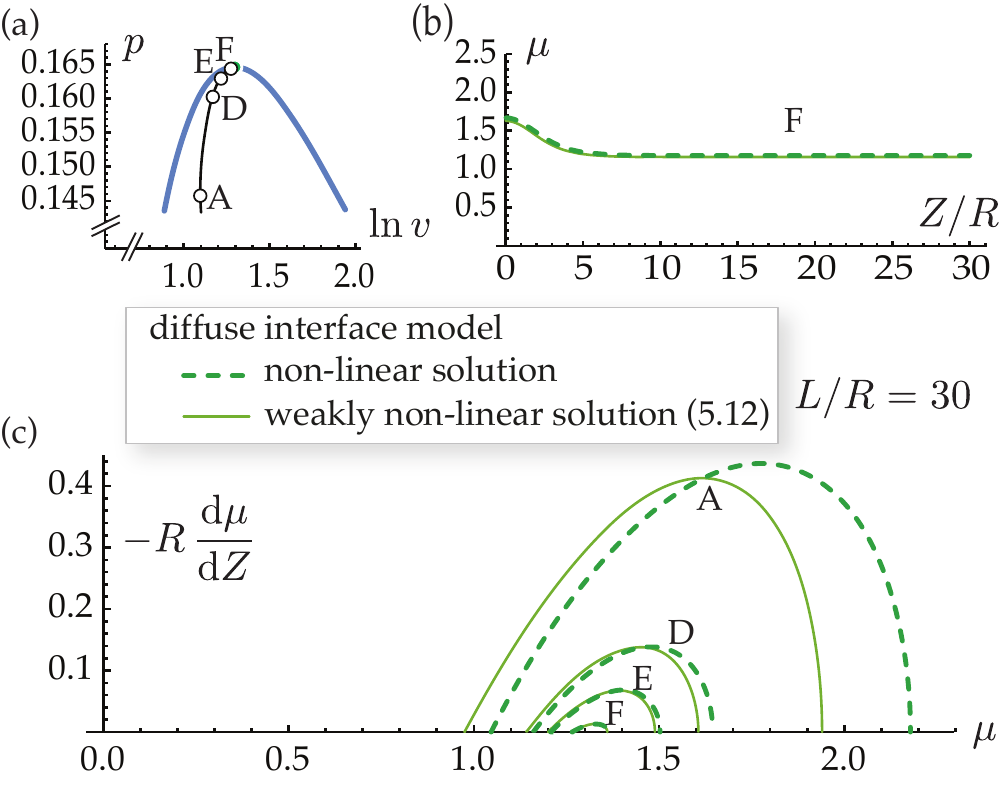}}
	\caption{Comparison of the analytical weakly non-linear
	solutions~(\ref{eq:weaklyNLSolutionSoliton}) and the numerical
	non-linear solutions of the diffuse interface model, for an
	aspect-ratio $L/R=30$.  The simulation domain represents one half
	of a balloon having a single bulge in the center.  (a)~Position of
	a few selected configurations in the phase diagram; point
	$\mathrm{A}$ is the same configuration as in
	figure~\ref{fig:fullnonlinear_solutions}.  (b)~Comparison close to
	the bifurcation point.  (c) Comparison for different bulging
	amplitudes in phase space: the weakly non-linear solution is
	asymptotically exact for small bulging amplitudes.}
	\protect\label{fig:soliton-test}
\end{figure}

In the right-hand side above, we can discard the terms that do not
depend on $\mu$, as well as the terms containing
$G_{0,\mu}^{\mathrm{C}}$ which cancels by
equation~(\ref{eq:dGdmuIsZero}), and that containing
$G_{0,\mu^{2}}^\mathrm{C}$ which cancels a the maximum pressure
$p_{\mathrm{C}}$.  Accordingly, the
energy~(\ref{eq:diffuseInterfaceModel-Announce}) can be approximated
as
\begin{equation}
	\mathcal{E} \approx \int_{-\infty}^{+\infty} \left[
	\mathrm{Cte}(p)
	+
	G_{0,p\mu}^\mathrm{C}\,(p-p_{\mathrm{C}})\,(\mu-\mu_{\mathrm{C}})
	+
	\frac{G_{0,\mu^{3}}^\mathrm{C}}{6}\,(\mu-\mu_{\mathrm{C}})^{3}
	+
	\frac{1}{2}\,B_{0}^{\mathrm{C}}\,\left(\frac{\mathrm{d}\mu}{\mathrm{d}Z}\right)^{2}
	\right]\,\mathrm{d}Z
	\textrm{.}
	\label{eq:energyExpansionSech2}
\end{equation}
The signs of the coefficients appearing in the integrand are
important: for our particular constitutive law and using the results
of Section~\ref{s:homsol_barmod}, their numerical value is
\begin{equation}
	G_{0,p\mu}^\mathrm{C} = \frac{\partial^{2} G_{0}}{\partial 
	p\,\partial \mu}(p_{\mathrm{C}},\mu_{\mathrm{C}}) = -9.366,
	\quad
	G_{0,\mu^{3}}^\mathrm{C} = \frac{\partial^{3} G_{0}}{\partial 
	\mu^{3}}(p_{\mathrm{C}},\mu_{\mathrm{C}}) = -3.413,\quad
	B_{0}^{\mathrm{C}}=B_{0}(p_{\mathrm{C}},\mu_{\mathrm{C}}) = 0.8956
	\nonumber
	\textrm{.}
\end{equation}
A balance argument on the three last term in the integrand above 
suggests the change of variable
\begin{subequations}
\begin{equation}
	\mu(Z) = \mu_{\mathrm{C}} + (p_{\mathrm{C}} - 
	p)^{1/2}\,\mu^{\dag}\,\overline{\mu}(\overline{Z})
	\label{eq:muRescaleSoliton}
\end{equation}
where
\begin{equation}
	\mu^{\dag} = \left(\frac{2\,|G_{0,p\mu}^\mathrm{C}|}{|
	G_{0,\mu^{3}}^\mathrm{C}
	|}\right)^{1/2},\quad
	\overline{Z} = 
	\left(
	\frac{2\,|G_{0,p\mu}^\mathrm{C}|\,|G_{0,\mu^{3}}^\mathrm{C}|}{
	(B_{0}^\mathrm{C})^{2}
	}\,
	(p_{\mathrm{C}}-p)
	\right)^{1/4}
	\, Z
	\textrm{.}
	\label{eq:ZbarSoliton}
\end{equation}
\end{subequations}
In terms of the rescaled variables, the energy
expansion~(\ref{eq:energyExpansionSech2}) writes
\begin{equation}
	\overline{\mathcal{E}} = \frac{1}{2}\int_{-\infty}^{+\infty} \left[
	\overline{\mu}(\overline{Z})-\frac{\overline{\mu}^{3}(\overline{Z})}{3}
	+ 
	\left(\frac{\mathrm{d}\overline{\mu}}{\mathrm{d}\overline{Z}}\right)^{2}
	\right] \,\mathrm{d}\overline{Z},
	\nonumber
\end{equation}
after dropping the term $G_{0}(p,\mu_{\mathrm{C}})$ that is
independent of $\mu$, and rescaling the energy using a numerical
constant.  The weakly non-linear solutions are the stationary points
$\overline{\mu}(\overline{Z})$ of this energy functional.  They can be
analyzed based on the analogy with a mass moving in a potential
$U(\overline{\mu}) =
-\frac{\overline{\mu}}{2}+\frac{\overline{\mu}^{3}}{6}$, when
$\overline{Z}$ is viewed as a time variable.  A first type of
solutions are those corresponding to the equilibria in the effective
potential $U(\overline{\mu})$, namely $\overline{\mu}(\overline{Z}) =
\pm 1$: they yield an expansion of the branch of homogeneous solutions
near the point of maximum pressure, as can be checked.  A second type
of solution corresponds to a soliton, \emph{i.e.}\ to a non-constant
but bounded solution; it can be derived by a quadrature method, using
the conservation of the total mechanical energy of the mass in the
effective potential.  The result is
\begin{equation}
	\overline{\mu}(\overline{Z}) = -1 + 
	\frac{3}{\cosh^{2}\frac{\overline{Z} - \overline{Z}_{0}}{2}}
	\textrm{.}
	\label{eq:weaklyNLSolutionSoliton}
\end{equation}
This solution represents a weakly localized bulge centered about
$\overline{Z}_{0}$.  It is identical to that derived
by Fu \textit{et al.}~\cite{Fu-Pearce-EtAl-Post-bifurcation-analysis-of-a-thin-walled-2008}
based on the full membrane model.  The soliton
solution~(\ref{eq:weaklyNLSolutionSoliton}) is plotted in
figure~\ref{fig:soliton-test}b--c, and compared to the non-linear
solution of the diffuse interface model: the weakly non-linear
solution and the full non-linear solution agree asymptotically close
to the bifurcation point, as expected.
Unlike the weakly non-linear solution from the previous section,
this solution captures the rapid localization of the bulge past the
bifurcation point: from~(\ref{eq:ZbarSoliton}), the width of the
interface is $Z\sim (p-p_{\mathrm{C}})^{-1/4}$ near bifurcation.

To derive the weakly non-linear solution, we have retained some terms
in the expansion of the energy~(\ref{eq:energyExpansionSech2}), and
omitted others such as $\frac{1}{2}\,G_{0,\mu
p^{2}}^\mathrm{C}\,(\mu-\mu_{\mathrm{C}})\,(p-p_{\mathrm{C}})^{2}$ or
$\frac{1}{2}\,B_{0,p}^\mathrm{C}\,(p-p_{\mathrm{C}})\,{\mu'}^{2}$.
This can be justified \emph{a posteriori}, based on the scaling laws
of the weakly non-linear solution: the scaling assumptions
$(\mu-\mu_{\mathrm{C}}) \sim (p_{\mathrm{C}}-p)^{1/2}$ and $Z\sim
(p_{\mathrm{C}}-p)^{-1/4}$ not only make the last three terms
in~(\ref{eq:energyExpansionSech2}) balanced (which is by design), they
also make the other terms negligible, as can be checked.  The
expansion~(\ref{eq:energyExpansionSech2})  is therefore consistent.

\section{Conclusion and final remarks}
\label{s:discussion}

We have proposed a diffuse interface model for the analysis of the
formation, localization and propagation of bulges in cylindrical
rubber balloons.  The model has been derived from the non-linear
membrane theory, and is asymptotically exact in the limit where the
strain gradient $\frac{\mathrm{d}\mu}{\mathrm{d}Z}$ is small compared
to $1/R$.  Analytical and numerical solutions to the diffuse interface
model have been obtained, showing good agreement with the predictions
of the original non-linear membrane model, both at the onset of
localization and for well-localized solutions: in practice, the
diffuse interface model remains accurate well beyond its domain of
strict mathematical validity, \emph{i.e.}\ even for relative large
gradients such as those found at the boundary between a bulged and a
non-bulged domain.  Hopefully, our work will shed new light on the
classical problem of bulging in elastic balloons, and will help
highlighting its tight connection with the theory of phase
transitions.

The model handles finite strain: only the strain \emph{gradients} are
assumed to be small.  The elastic response of the material under
finite strain gets reflected into the diffuse interface model through
the non-linear potential $G_{0}(p,\mu)$ and the non-linear strain
gradient modulus $B_{0}(p,\mu)$.  Thanks to this feature, the
predictions of the model are exact as far as linear bifurcation
analyses are concerned, and remain accurate even well into the
post-bifurcation regime.  By contrast, expansion methods underlying
bifurcation analyses typically assume that the solution is close to a
homogeneous solution, and have a much narrower domain of
applicability.

A simple expression for the coefficient $B_{0}(p,\mu)$ of the strain
gradient term has been established, see~(\ref{eq:Bexpression}).
Remarkably, this coefficient is directly proportional to the
pre-stress $\boldsymbol{\Sigma}_0$ in the homogeneous solution,
see~(\ref{eq:E-tmp1}).  The same observation has been made concerning
the strain-gradient model applicable to necking in a hyperelastic
cylinder~\cite{BAJH}.  In both cases, the contribution of the strain
gradient to the elastic energy occurs through a `geometric rigidity
effect' --- the vibration of a string under tension is another example
of this geometric rigidity effect, whereby the pre-stress brings in an
effective elastic stiffness.

The consistency of our results with those of Audoly \&
Hutchinson~\cite{BAJH} for a solid cylinder can be checked as follows.
In~(\ref{eq:Efinal}), we have derived the contribution of the strain
gradient to the energy of a balloon using dimensionless quantities.
In terms of the original (non-scaled) quantities, it can be rewritten
as $\frac{1}{2}\int_{0}^{L}B_{0}^{*}\,\mu'^{2}(Z)\,\mathrm{d}Z$, where
the non-scaled modulus is found from~(\ref{eq:Bexpression}) as
$B_{0}^{*} = (2\pi\,R\,H)\,R^{2}\,\frac{1}{\lambda}\frac{\partial
{w}_{0}^{*}}{\partial \lambda}$, after restoring the initial area that
had been scaled out.  Identifying $I=(2\pi\,R\,H)\,R^{2}$ as the
geometric moment of inertia of the annular cross-section in reference
configuration, the contribution of the strain gradient to the energy
can therefore be written as
$\frac{1}{2}\int_{0}^{L}I\,\frac{1}{\lambda}\frac{\partial
{w}_{0}^{*}}{\partial \lambda}\,\mu'^{2}(Z)\,\mathrm{d}Z$ in an
axisymmetric membrane.  In fact, this holds for a solid cylinder as
well, as can be checked by combining equations~2.28, 2.13 and~2.8
from~\cite{BAJH}.

In the future, the following extensions of the model could be
considered.  While we have only been concerned with bifurcations in
this paper, one could analyze the stability of the solutions based on
the diffuse interface model as well: presumably, this would confirm
the stability results obtained previously from the non-linear membrane
model.  Different boundary conditions than the natural boundary
conditions could be used at the ends of the balloons: a hard plug can
be enforced by the boundary condition $\mu=1$, while a rounded elastic
cap can be prescribed through a non-cylindrical reference
configuration $R(Z)$; if, however, the prescribed initial profile $R$
varies quickly with $Z$, there may be a conflict with our asymptotic
procedure that assumes slow variations, and the diffuse interface
model may not be able to account accurately for such boundary
conditions.  A coupling with an electrical field could also be
introduced, as is relevant to the actuation of balloons made of
dielectric elastomers~\cite{
Lu-Suo-Large-conversion-of-energy-2012,%
Lu-An-EtAl-Electro-mechanical-coupling-bifurcation-2015,%
An-Wang-EtAl-Experimental-investigation-of-the-electromechanical-2015};
we believe that a natural extension of our asymptotic expansion can be
derived for dielectric balloons.
In future work, we also hope to apply our asymptotic reduction method
to other localization phenomena occurring in slender structures, such
as the Plateau-Rayleigh instability in soft elastic rods with surface
tension~\cite{Xuan-Biggins-Plateau-Rayleigh-instability-in-solids-2017},
and the localized kinks in tape 
springs~\cite{Seffen-Pellegrino-Deployment-dynamics-of-tape-1999,%
Seffen-You-EtAl-Folding-and-deployment-of-curved-2000}.

\textit{We would like to thank John Hutchinson for drawing our attention
to the absence of a strain gradient model for the analysis of bulging
in balloons, and for providing useful feedback on the manuscript.}

\appendix

\section{Elimination of $\lambda(Z)$ from the diffuse interface model}
\label{B:Redmodel}

This appendix provides details on the elimination of the unknown
$\lambda(Z)$ from the intermediate form~(\ref{eq:E-tmp1}) of the
energy, leading to the final form~(\ref{subeqs:EfinalAndB0}).

As the approximation of the energy~(\ref{eq:E-tmp1}) contains no derivative of the axial stretch $\lambda$, optimizing with respect with the $\lambda$ variable
yields an algebraic ({\emph{i.e.}}~non-differential) problem for $\lambda
(Z)$
\begin{equation}
\frac{\partial g_0}{\partial \lambda} (p, \lambda (Z), \mu (Z)) +
\frac{\partial \boldsymbol{\Sigma}_0}{\partial \lambda} (\lambda (Z), \mu (Z))
\cdot \mathbf{E}_1 (\mu' (Z)) +\mathcal{O} (\varepsilon^4) = 0.
\label{eq:nonHomAxialEquil}
\end{equation}
The first term is of order $\varepsilon^0 = 1$, the following term is of order
$\varepsilon^2$ by~(\ref{eq:mainScalings}), and we have omitted terms of order
$\varepsilon^4$ and higher.

Let us solve~(\ref{eq:nonHomAxialEquil}) order by order for $\lambda$.
At dominant order $\varepsilon^0 = 1$, $\lambda (Z)$ is a solution of
$\frac{\partial g_0}{\partial \lambda} (p, \lambda, \mu (Z)) = 0$,
which we identify as the equilibrium condition~(\ref{eq:g0EquiAxial})
applicable to homogeneous solutions.  In~(\ref{eq:lambda0def}), the
solution to this equation has be=en introduced as $\lambda (Z) =
\lambda_0 (p, \mu (Z))$, {\emph{i.e.}}
\[ \lambda (Z) = \lambda_0 (p, \mu (Z)) +\mathcal{O} (\varepsilon^2) . \]
To order $\varepsilon^{2}$, the solution of
equation~(\ref{eq:nonHomAxialEquil}) is found as
\begin{equation}
\lambda (Z) = \lambda_0 (p, \mu (Z)) + \lambda_{[2]} (Z) +\mathcal{O}
(\varepsilon^4), \label{eq:lambdaToOrderEps2}
\end{equation}
where $\lambda_{[2]} =-\left( \frac{\partial
\boldsymbol{\Sigma}_0}{\partial \lambda} \cdot \mathbf{E}_1 (\mu')
\right) / \frac{\partial^2 g_0}{\partial \lambda^2}$ is a correction
of order $\varepsilon^2$, arising from the strain correction
$\mathbf{E}_1 (\mu') =\mathcal{O} (\varepsilon^2)$ (the expression
of $\lambda_{[2]}$ is given for the sake of completeness but is not
needed in the following).

Inserting the expansion~(\ref{eq:lambdaToOrderEps2}) for $\lambda (Z)$
into~(\ref{eq:E-tmp1}) and expanding in series, we find
\[ \mathcal{E}= \int \left[ G_0 (p, \mu) + \frac{\partial g_0}{\partial
	\lambda} (p, \lambda_0, \mu) \,\lambda_{[2]} \right] \,\mathrm{d} Z + \int
\boldsymbol{\Sigma}_0 (\lambda_0, \mu) \cdot \mathbf{E}_1 (\mu')\, \mathrm{d} Z
+\mathcal{O} (L \,\varepsilon^4), \]
where we use $\lambda_0$ as a shorthand for $\lambda_0 (p, \mu (Z))$.
As the second term in the bracket vanishes by~(\ref{eq:lambda0def}),
we are left with
\[ \mathcal{E}= \int G_0 (p, \mu)\, \mathrm{d} Z + \int \boldsymbol{\Sigma}_0
(\lambda_0, \mu) \cdot \mathbf{E}_1 (\mu') \,\mathrm{d} Z +\mathcal{O} (L\,
\varepsilon^4) . \]
Inserting the explicit expressions of $\boldsymbol{\Sigma}_0$ and 
$\mathbf{E}_1$ from~(\ref{eq:sigma0}) and~(\ref{eq:E1}), one 
obtains the expressions announced in~(\ref{subeqs:EfinalAndB0}).

\bibliographystyle{elsarticle-num-names}


\begin{thebibliography}{45}
	\providecommand{\natexlab}[1]{#1}
	\providecommand{\url}[1]{\texttt{#1}}
	\providecommand{\urlprefix}{URL }
	\expandafter\ifx\csname urlstyle\endcsname\relax
	\providecommand{\doi}[1]{doi:\discretionary{}{}{}#1}\else
	\providecommand{\doi}[1]{doi:\discretionary{}{}{}\begingroup
		\urlstyle{rm}\url{#1}\endgroup}\fi
	\providecommand{\bibinfo}[2]{#2}
	
	\bibitem[{Kyriakides and
		Chang(1991)}]{Kyriakides-Chang-The-initiation-and-propagation-of-a-localized-1991}
	\bibinfo{author}{S.~Kyriakides}, \bibinfo{author}{Y.-C. Chang},
	\bibinfo{title}{The initiation and propagation of a localized instability in
		an inflated elastic tube}, \bibinfo{journal}{International Journal of Solids
		and Structures} \bibinfo{volume}{27}~(\bibinfo{number}{9})
	(\bibinfo{year}{1991}) \bibinfo{pages}{1085--1111}.
	
	\bibitem[{Chater and
		Hutchinson(1984)}]{Chater-Hutchinson-On-the-propagation-of-bulges-and-buckles-1984}
	\bibinfo{author}{E.~Chater}, \bibinfo{author}{J.~W. Hutchinson},
	\bibinfo{title}{On the propagation of bulges and buckles},
	\bibinfo{journal}{Journal of Applied Mechanics}
	\bibinfo{volume}{51}~(\bibinfo{number}{2}) (\bibinfo{year}{1984})
	\bibinfo{pages}{269--277}.
	
	\bibitem[{Ericksen(1975)}]{Ericksen75}
	\bibinfo{author}{J.~L. Ericksen}, \bibinfo{title}{Equilirbium of bars},
	\bibinfo{journal}{Journal of Elasticity} \bibinfo{volume}{5}
	(\bibinfo{year}{1975}) \bibinfo{pages}{191}.
	
	\bibitem[{Bhattacharya(2004)}]{Bhattacharya-Oxford-Series-on-Materials-2004}
	\bibinfo{author}{S.~Bhattacharya}, \bibinfo{title}{Microstructure of
		martensite}, Oxford Series on materials modelling, \bibinfo{publisher}{Oxford
		University Press}, \bibinfo{year}{2004}.
	
	\bibitem[{Fu and Freidin(2004)}]{FuFreidin04}
	\bibinfo{author}{Y.~Fu}, \bibinfo{author}{A.~Freidin},
	\bibinfo{title}{Characterization and stability of two-phase
		piecewise-homogeneous deformations}, \bibinfo{journal}{Proc. R. Soc. A}
	\bibinfo{volume}{460} (\bibinfo{year}{2004}) \bibinfo{pages}{2065}.
	
	\bibitem[{Bridgman(1952)}]{Bridgman-Studies-in-large-plastic-1952}
	\bibinfo{author}{P.~W. Bridgman}, \bibinfo{title}{Studies in large plastic flow
		and fracture}, Metallurgy and metallurgical engineering series,
	\bibinfo{publisher}{Mc{\-}Graw-Hill}, \bibinfo{address}{New York},
	\bibinfo{year}{1952}.
	
	\bibitem[{Barenblatt(1974)}]{Barenblatt-Neck-propagation-in-polymers-1974}
	\bibinfo{author}{G.~I. Barenblatt}, \bibinfo{title}{Neck propagation in
		polymers}, \bibinfo{journal}{Rheologica Acta} \bibinfo{volume}{13}
	(\bibinfo{year}{1974}) \bibinfo{pages}{924--933}.
	
	\bibitem[{Hutchinson and
		Miles(1974)}]{Hutchinson-Miles-Bifurcation-analysis-of-the-onset-1974}
	\bibinfo{author}{J.~W. Hutchinson}, \bibinfo{author}{J.~P. Miles},
	\bibinfo{title}{Bifurcation analysis of the onset of necking in an
		elastic/plastic cylinder under uniaxial tension}, \bibinfo{journal}{Journal
		of the Mechanics and Physics of Solids} \bibinfo{volume}{22}
	(\bibinfo{year}{1974}) \bibinfo{pages}{61--71}.
	
	\bibitem[{Wadee et~al.(2004)Wadee, Hunt, and
		Peletier}]{Wadee-Hunt-EtAl-Kink-band-instabilities-2004}
	\bibinfo{author}{M.~A. Wadee}, \bibinfo{author}{G.~W. Hunt},
	\bibinfo{author}{M.~A. Peletier}, \bibinfo{title}{Kink band instabilities in
		layered structures}, \bibinfo{journal}{Journal of the Mechanics and Physics
		of Solids} \bibinfo{volume}{52} (\bibinfo{year}{2004})
	\bibinfo{pages}{1071--1091}.
	
	\bibitem[{Fu and Zhang(2006)}]{FuZhang06}
	\bibinfo{author}{Y.~Fu}, \bibinfo{author}{Y.~Zhang},
	\bibinfo{title}{Continuum-mechanical modelling of kink-band formation in
		fibre-reinforced composites}, \bibinfo{journal}{International Journal of
		Solids and Structures} \bibinfo{volume}{43} (\bibinfo{year}{2006})
	\bibinfo{pages}{3306}.
	
	\bibitem[{Power and
		Kyriakides(1994)}]{Power-Kyriakides-Localization-and-propagation-of-instabilities-1994}
	\bibinfo{author}{T.~L. Power}, \bibinfo{author}{S.~Kyriakides},
	\bibinfo{title}{Localization and propagation of instabilities in long shallow
		panels under external pressure}, \bibinfo{journal}{Journal of Applied
		Mechanics} \bibinfo{volume}{61} (\bibinfo{year}{1994})
	\bibinfo{pages}{755--763}.
	
	\bibitem[{Seffen and
		Pellegrino(1999)}]{Seffen-Pellegrino-Deployment-dynamics-of-tape-1999}
	\bibinfo{author}{K.~A. Seffen}, \bibinfo{author}{S.~Pellegrino},
	\bibinfo{title}{Deployment dynamics of tape springs},
	\bibinfo{journal}{Proceedings of the Royal Society of London. Series A:
		Mathematical, Physical and Engineering Sciences}
	\bibinfo{volume}{455}~(\bibinfo{number}{1983}) (\bibinfo{year}{1999})
	\bibinfo{pages}{1003--1048}.
	
	\bibitem[{Seffen et~al.(2000)Seffen, You, and
		Pellegrino}]{Seffen-You-EtAl-Folding-and-deployment-of-curved-2000}
	\bibinfo{author}{K.~A. Seffen}, \bibinfo{author}{Z.~You},
	\bibinfo{author}{S.~Pellegrino}, \bibinfo{title}{Folding and deployment of
		curved tape springs}, \bibinfo{journal}{International Journal of Mechanical
		Sciences} \bibinfo{volume}{42}~(\bibinfo{number}{10}) (\bibinfo{year}{2000})
	\bibinfo{pages}{2055--2073}.
	
	\bibitem[{Triantafyllidis and
		Bardenhagen(1996)}]{Triantafyllidis-Bardenhagen-The-influence-of-scale-size-1996}
	\bibinfo{author}{N.~Triantafyllidis}, \bibinfo{author}{S.~Bardenhagen},
	\bibinfo{title}{The influence of scale size on the stability of periodic
		solids and the role of associated higher order gradient continuum models},
	\bibinfo{journal}{Journal of the Mechanics and Physics of Solids}
	\bibinfo{volume}{44} (\bibinfo{year}{1996}) \bibinfo{pages}{1891--1928}.
	
	\bibitem[{Knowles and Sternberg(1978)}]{KnowlesSternberg}
	\bibinfo{author}{J.~Knowles}, \bibinfo{author}{E.~Sternberg},
	\bibinfo{title}{On the failure ot ellipticity and the emergence of
		discontinuous deformation gradients in plane finite elastostatics},
	\bibinfo{journal}{journal of Elasticity} \bibinfo{volume}{4}
	(\bibinfo{year}{1978}) \bibinfo{pages}{329}.
	
	\bibitem[{Triantafyllidis and Aifantis(1986)}]{TriantaAifan}
	\bibinfo{author}{N.~Triantafyllidis}, \bibinfo{author}{E.~C. Aifantis},
	\bibinfo{title}{A gradient approach to localization of deformation. {I}.
		Hyperelastic materials}, \bibinfo{journal}{Journal of Elasticity}
	\bibinfo{volume}{16} (\bibinfo{year}{1986}) \bibinfo{pages}{225--237}.
	
	\bibitem[{Triantafyllidis and
		Bardenhagen(1993)}]{Triantafyllidis-Bardenhagen-On-higher-order-gradient-1993}
	\bibinfo{author}{N.~Triantafyllidis}, \bibinfo{author}{S.~Bardenhagen},
	\bibinfo{title}{On higher order gradient continuum theories in {1-D}
		nonlinear elasticity. {Derivation} from and comparison to the corresponding
		discrete models}, \bibinfo{journal}{Journal of Elasticity}
	\bibinfo{volume}{33}~(\bibinfo{number}{3}) (\bibinfo{year}{1993})
	\bibinfo{pages}{259--293}.
	
	\bibitem[{Bardenhagen and Triantafyllidis(1994)}]{BardenTrianta94}
	\bibinfo{author}{S.~Bardenhagen}, \bibinfo{author}{N.~Triantafyllidis},
	\bibinfo{title}{Derivation of higher order gradient continuum theories in
		2,3-d non-linear elasticity from periodic lattice models},
	\bibinfo{journal}{J. Mech. Phys. Solids} \bibinfo{volume}{42}
	(\bibinfo{year}{1994}) \bibinfo{pages}{111--139}.
	
	\bibitem[{Abdoul-Anziz and
		Seppecher(2018)}]{Abdoul-Anziz-Seppecher-Strain-gradient-and-generalized-2018}
	\bibinfo{author}{H.~Abdoul-Anziz}, \bibinfo{author}{P.~Seppecher},
	\bibinfo{title}{Strain gradient and generalized continua obtained by
		homogenizing frame lattices}, \bibinfo{journal}{Mathematics and Mechanics of
		Complex Systems} .
	
	\bibitem[{Bacigalupo et~al.(2017)Bacigalupo, Paggi, Dal~Corso, and
		Bigoni}]{Bacigalupo-Paggi-EtAl-Identification-of-higher-order-continua-2017}
	\bibinfo{author}{A.~Bacigalupo}, \bibinfo{author}{M.~Paggi},
	\bibinfo{author}{F.~Dal~Corso}, \bibinfo{author}{D.~Bigoni},
	\bibinfo{title}{Identification of higher-order continua equivalent to a
		{Cauchy} elastic composite}, \bibinfo{journal}{Mechanics Research
		Communications} .
	
	\bibitem[{Mielke(1991)}]{Mielke-Hamiltonian-and-Lagrangian-flows-1991}
	\bibinfo{author}{A.~Mielke}, \bibinfo{title}{{Hamiltonian} and {Lagrangian}
		flows on center manifolds, with application to elliptic variational
		problems}, vol. \bibinfo{volume}{1489} of \emph{\bibinfo{series}{Lecture
			notes in mathematics}}, \bibinfo{publisher}{Springer-Verlag},
	\bibinfo{address}{Berlin}, \bibinfo{year}{1991}.
	
	\bibitem[{Audoly and Hutchinson(2016)}]{BAJH}
	\bibinfo{author}{B.~Audoly}, \bibinfo{author}{J.~W. Hutchinson},
	\bibinfo{title}{Analysis of necking based on a one-dimensional model},
	\bibinfo{journal}{Journal of Mechanics Physics of Solids}
	\bibinfo{volume}{97} (\bibinfo{year}{2016}) \bibinfo{pages}{68--91}.
	
	\bibitem[{Kyriakides and Chang(1990)}]{KyriaChang90}
	\bibinfo{author}{S.~Kyriakides}, \bibinfo{author}{Y.-C. Chang},
	\bibinfo{title}{On the inflation of a long elastic tube in the presence of
		axial load}, \bibinfo{journal}{International Journal of Solids and
		Structures} \bibinfo{volume}{26} (\bibinfo{year}{1990}) \bibinfo{pages}{975}.
	
	\bibitem[{van~der Waals(1894)}]{vdwaals94}
	\bibinfo{author}{J.~D. van~der Waals}, \bibinfo{title}{{Thermodynamische
			Theorie der Kapillarit{\"a}t unter Voraussetzung stetiger
			Dichte{\"a}nderung}.}, \bibinfo{journal}{Z. Phys. Chem.} \bibinfo{volume}{13}
	(\bibinfo{year}{1894}) \bibinfo{pages}{657--725}.
	
	\bibitem[{M\"{u}ller and
		Strehlow(2004)}]{Muller-Strehlow-Rubber-and-Rubber-Balloons-2004}
	\bibinfo{author}{A.~M\"{u}ller}, \bibinfo{author}{P.~Strehlow},
	\bibinfo{title}{Rubber and Rubber Balloons}, \bibinfo{publisher}{Springer},
	\bibinfo{year}{2004}.
	
	\bibitem[{Corneliussen and Shield(1961)}]{CorneliussenShield61}
	\bibinfo{author}{A.~Corneliussen}, \bibinfo{author}{R.~T. Shield},
	\bibinfo{title}{Finite Deformation of Elastic Membranes with Application to
		the Stability of an Inflated and Extended Tube}, \bibinfo{journal}{Arch.
		Rational Mech. Anal.} \bibinfo{volume}{7} (\bibinfo{year}{1961})
	\bibinfo{pages}{273}.
	
	\bibitem[{Shield(1972)}]{Shield72}
	\bibinfo{author}{R.~T. Shield}, \bibinfo{title}{On the stability of finitely
		deformed elastic membranes, Part {II}: stability of inflated cylindrical and
		spherical membranes.}, \bibinfo{journal}{J. Appl. Math. Phys.}
	\bibinfo{volume}{23}.
	
	\bibitem[{Haughton and Odgen(1979)}]{HaughtonOgden79}
	\bibinfo{author}{D.~M. Haughton}, \bibinfo{author}{R.~W. Odgen},
	\bibinfo{title}{Bifurcation of inflated circular cylinders of elastic
		material under axial loading, part i: membrane theory for thin-walled tubes},
	\bibinfo{journal}{Journal of the Mechanics and Physics of Solids}
	\bibinfo{volume}{27} (\bibinfo{year}{1979}) \bibinfo{pages}{179}.
	
	\bibitem[{Yin(1977)}]{Yin-Non-uniform-inflation-of-a-cylindrical-1977}
	\bibinfo{author}{W.-L. Yin}, \bibinfo{title}{Non-uniform inflation of a
		cylindrical elastic membrane and direct determination of the strain energy
		function}, \bibinfo{journal}{Journal of Elasticity}
	\bibinfo{volume}{7}~(\bibinfo{number}{3}) (\bibinfo{year}{1977})
	\bibinfo{pages}{265--282}.
	
	\bibitem[{Chen(1997)}]{chenbulges}
	\bibinfo{author}{Y.-C. Chen}, \bibinfo{title}{Stability and bifurcation of
		finite deformations of elastic cylindrical membranes, part {I}: Stability
		analysis}, \bibinfo{journal}{International Journal of Solids and Structures}
	\bibinfo{volume}{34} (\bibinfo{year}{1997}) \bibinfo{pages}{1735}.
	
	\bibitem[{Fu et~al.(2008)Fu, Pearce, and
		Liu}]{Fu-Pearce-EtAl-Post-bifurcation-analysis-of-a-thin-walled-2008}
	\bibinfo{author}{Y.~B. Fu}, \bibinfo{author}{S.~P. Pearce},
	\bibinfo{author}{K.~K. Liu}, \bibinfo{title}{Post-bifurcation analysis of a
		thin-walled hyperelastic tube under inflation},
	\bibinfo{journal}{International Journal of Non-Linear Mechanics}
	\bibinfo{volume}{43}~(\bibinfo{number}{8}) (\bibinfo{year}{2008})
	\bibinfo{pages}{697--706}.
	
	\bibitem[{Fu and Xie(2010)}]{Fu-Xie-Stability-of-localized-bulging-2010}
	\bibinfo{author}{Y.~B. Fu}, \bibinfo{author}{Y.~X. Xie},
	\bibinfo{title}{Stability of localized bulging in inflated membrane tubes
		under volume control}, \bibinfo{journal}{International Journal of Engineering
		Science} \bibinfo{volume}{48} (\bibinfo{year}{2010})
	\bibinfo{pages}{1242--1252}.
	
	\bibitem[{Pearce and
		Fu(2010)}]{Pearce-Fu-Characterization-and-stability-of-localized-2010}
	\bibinfo{author}{S.~P. Pearce}, \bibinfo{author}{Y.~B. Fu},
	\bibinfo{title}{Characterization and stability of localized bulging/necking
		in inflated membrane tubes}, \bibinfo{journal}{IMA Journal of Applied
		Mathematics} \bibinfo{volume}{75} (\bibinfo{year}{2010})
	\bibinfo{pages}{581--602}.
	
	\bibitem[{Fu and Xie(2012)}]{Fu-Effects-of-imperfections-on-localized-2012}
	\bibinfo{author}{Y.~B. Fu}, \bibinfo{author}{Y.~X. Xie},
	\bibinfo{title}{Effects of imperfections on localized bulging in inflated
		membrane tubes}, \bibinfo{journal}{Philosophical Transactions of the Royal
		Society A: Mathematical, Physical and Engineering Sciences}
	\bibinfo{volume}{370} (\bibinfo{year}{2012}) \bibinfo{pages}{1896--1911}.
	
	\bibitem[{Coleman and Newman(1988)}]{ColemanNewman}
	\bibinfo{author}{B.~D. Coleman}, \bibinfo{author}{D.~C. Newman},
	\bibinfo{title}{On the Rheology of Cold Drawing. I. Elastic Materials},
	\bibinfo{journal}{Journal of Polymer Science: Part B: Polymer Physics}
	\bibinfo{volume}{26} (\bibinfo{year}{1988}) \bibinfo{pages}{1801}.
	
	\bibitem[{Dai and Cai(2006)}]{CaiDai06P1}
	\bibinfo{author}{H.-H. Dai}, \bibinfo{author}{Z.~Cai}, \bibinfo{title}{Phase
		transitions in a slender cylinder composed of an incompressible elastic
		material. {I. Asymptotic} model equation}, \bibinfo{journal}{Proc. R. Soc. A}
	\bibinfo{volume}{462} (\bibinfo{year}{2006}) \bibinfo{pages}{75--95}.
	
	\bibitem[{Dai and
		Wang(2009)}]{Dai-Wang-An-analytical-study-on-the-geometrical-2009}
	\bibinfo{author}{H.-H. Dai}, \bibinfo{author}{J.~Wang}, \bibinfo{title}{An
		analytical study on the geometrical size effect on phase transitions in a
		slender compressible hyperelastic cylinder}, \bibinfo{journal}{International
		Journal of Non-Linear Mechanics} \bibinfo{volume}{44} (\bibinfo{year}{2009})
	\bibinfo{pages}{219--229}.
	
	\bibitem[{Ogden(1972)}]{Ogden-Large-deformation-isotropic-1972}
	\bibinfo{author}{R.~W. Ogden}, \bibinfo{title}{Large deformation isotropic
		elasticity-on the correlation of theory and experiment for incompressible
		rubber-like solids}, \bibinfo{journal}{Proceedings of the Royal Society A:
		Mathematical, Physical and Engineering Science} \bibinfo{volume}{326}
	(\bibinfo{year}{1972}) \bibinfo{pages}{565--584}.
	
	\bibitem[{Consid{\`e}re(1885)}]{Considere-Memoire-sur-lemploi-du-fer-et-de-lacier-1885}
	\bibinfo{author}{A.~Consid{\`e}re}, \bibinfo{title}{M{\'e}moire sur l'emploi du
		fer et de l'acier dans les constructions}, \bibinfo{journal}{Annales des
		Ponts et Chauss{\'e}es, S{\'e}rie 6} \bibinfo{volume}{9}
	(\bibinfo{year}{1885}) \bibinfo{pages}{574--775}.
	
	\bibitem[{Doedel et~al.(2007)Doedel, Champneys, Fairgrieve, Kuznetsov,
		Sandstede, and
		Wang}]{Doedel-Champneys-EtAl-AUTO-07p:-continuation-and-bifurcation-2007}
	\bibinfo{author}{E.~J. Doedel}, \bibinfo{author}{A.~R. Champneys},
	\bibinfo{author}{T.~F. Fairgrieve}, \bibinfo{author}{Y.~A. Kuznetsov},
	\bibinfo{author}{B.~Sandstede}, \bibinfo{author}{X.~J. Wang},
	\bibinfo{title}{{AUTO}-07p: continuation and bifurcation software for
		ordinary differential equations}, \bibinfo{howpublished}{See
		http://indy.cs.concordia.ca/auto/}, \bibinfo{year}{2007}.
	
	\bibitem[{Koiter(1965)}]{Koiter-On-the-stability-of-elastic-equilibrium-1965}
	\bibinfo{author}{W.~T. Koiter}, \bibinfo{title}{On the stability of elastic
		equilibrium}, Ph.D. thesis, \bibinfo{school}{Delft},
	\bibinfo{address}{Holland}, \bibinfo{year}{1965}.
	
	\bibitem[{Lu and Suo(2012)}]{Lu-Suo-Large-conversion-of-energy-2012}
	\bibinfo{author}{T.-Q. Lu}, \bibinfo{author}{Z.~Suo}, \bibinfo{title}{Large
		conversion of energy in dielectric elastomers by electromechanical phase
		transition}, \bibinfo{journal}{Acta Mechanica sinica}
	\bibinfo{volume}{28}~(\bibinfo{number}{4}) (\bibinfo{year}{2012})
	\bibinfo{pages}{1106--1114}.
	
	\bibitem[{Lu et~al.(2015)Lu, An, Li, Yuan, and
		Wang}]{Lu-An-EtAl-Electro-mechanical-coupling-bifurcation-2015}
	\bibinfo{author}{T.~Lu}, \bibinfo{author}{L.~An}, \bibinfo{author}{J.~Li},
	\bibinfo{author}{C.~Yuan}, \bibinfo{author}{T.~J. Wang},
	\bibinfo{title}{Electro-mechanical coupling bifurcation and bulging
		propagation in a cylindrical dielectric elastomer tube},
	\bibinfo{journal}{Journal of the Mechanics and Physics of Solids}
	\bibinfo{volume}{85} (\bibinfo{year}{2015}) \bibinfo{pages}{160--175}.
	
	\bibitem[{An et~al.(2015)An, Wang, Cheng, Lu, and
		Wang}]{An-Wang-EtAl-Experimental-investigation-of-the-electromechanical-2015}
	\bibinfo{author}{L.~An}, \bibinfo{author}{F.~Wang}, \bibinfo{author}{S.~Cheng},
	\bibinfo{author}{T.~Lu}, \bibinfo{author}{T.~J. Wang},
	\bibinfo{title}{Experimental investigation of the electromechanical phase
		transition in a dielectric elastomer tube}, \bibinfo{journal}{Smart Materials
		and Structures} \bibinfo{volume}{24}~(\bibinfo{number}{3})
	(\bibinfo{year}{2015}) \bibinfo{pages}{035006}.
	
	\bibitem[{Xuan and
		Biggins(2017)}]{Xuan-Biggins-Plateau-Rayleigh-instability-in-solids-2017}
	\bibinfo{author}{C.~Xuan}, \bibinfo{author}{J.~Biggins},
	\bibinfo{title}{{Plateau-Rayleigh} instability in solids is a simple phase
		separation}, \bibinfo{journal}{Physical Reivew E} \bibinfo{volume}{95}
	(\bibinfo{year}{2017}) \bibinfo{pages}{053106}.
	
\end{thebibliography}


\end{document}